\documentclass[useAMS,usenatbib,usegraphicx]{mn2e}
\usepackage{aas_macros}

\title[{\it HST} Survey for EROs]{An {\it HST}\thanks{Based on observations from
    the NASA/ESA {\it Hubble Space Telescope} obtained from the
    ESO/ST-ECF Science Archive Facility.} Morphological Survey of Faint
  EROs} \author[D.\,G.\ Gilbank et al.]  {David G.\ 
  Gilbank$^1$\thanks{Email: D.G.Gilbank@durham.ac.uk},
  Ian Smail$^1$, R.\,J.\ Ivison$^2$ and C.\ Packham$^3$\\
  $^1$Institute for Computational Cosmology, University of Durham,
  South Road, Durham,
  DH1 3LE\\
  $^2$Astronomy Technology Centre, Royal Observatory, Blackford Hill,
  Edinburgh, EH9 3HJ\\
  $^3$Department of Astronomy, University of Florida, 211 SSRB,
  Gainesville, Florida 32611, USA} \date{\today}

\def\gsim{\mathrel{\raise0.35ex\hbox{$\scriptstyle >$}\kern-0.6em
\lower0.40ex\hbox{{$\scriptstyle \sim$}}}}
\def\lsim{\mathrel{\raise0.35ex\hbox{$\scriptstyle <$}\kern-0.6em
\lower0.40ex\hbox{{$\scriptstyle \sim$}}}}

\pagerange{\pageref{firstpage}--\pageref{lastpage}} \pubyear{2003}

\def\ks {{$K_{\rm s}$}}
\def\j {{$J$}}
\def\sex {{\sc SExtractor}}
\def\i814 {{$I_{814}$}}
\def\yt {{YT03}}

\begin{document}

\label{firstpage}

\maketitle

\begin{abstract}
  We present the results from a survey for Extremely Red Objects (EROs)
  in deep, high resolution optical images taken from the {\it Hubble
    Space Telescope} ({\it HST}) Medium Deep Survey. We have surveyed
  35 deep F814W {\it HST}/WFPC2 fields in the near-infrared to a
  typical depth of $K\gsim 20$.  ~From a total area of 206 arcmin$^2$
  and to a limit of $K=20.0$ we identify 224 EROs ($(1.14\pm0.08)$
  arcmin$^{-2}$) with $(I_{814}-K)\ge4.0$ and 83 ($(0.41\pm0.05)$
  arcmin$^{-2}$) with $(I_{814}-K)\ge5.0$.  We find that the slope of
  the number counts of the $(I_{814}-K)\ge4.0$ EROs flattens beyond
  $K\sim 19$, in line with results from previous surveys, and the
  typical colours of the EROs become redder beyond the break magnitude.
  We morphologically classify our ERO sample using visual and
  quantitative schemes and find that 35\% of our sample exhibit clear
  disk components, 15\% are disturbed or irregular, a further 30\% are
  either spheroidal or compact and the remaining 20\% are
  unclassifiable. Using a quantitative measure of morphology, we find
  that the ERO morphological distribution evolves across the break in
  their counts, such that low concentration (disk-like) galaxies
  decline. We relate the morphological and colour information for our
  EROs and conclude that those EROs morphologically classified as
  bulges do indeed possess SEDs consistent with passive stellar
  populations; while EROs with dusty star-forming SEDs are mostly
  associated with disk-like and peculiar galaxies.  However, $\sim
  30\%$ of disk EROs reside in the passive region of I/J/K
  colour-colour space.  These could be either genuinely passive
  systems, lower redshift contaminants to the high-z ERO population, or
  systems with composite star-forming and passive SEDs.  We use
  photometric redshifts for our high S/N multicolour photometry and
  derive redshift distributions in good agreement with spectroscopic
  work of somewhat brighter ERO examples.
\end{abstract}


\section{Introduction}

Broadband photometric selection is an increasingly common technique to
identify high redshift galaxies
\citep{1999ApJ...519....1S,2001ApJ...558L..87T,2003ApJ...587L..79F}.
One particularly powerful combination is to use optical and
near-infrared passbands, typically $R$ or $I$ and $K$, to identify
galaxies with a strong decline in their spectral energy distributions
(SED) at wavelengths around 1$\mu$m.  These sources display extremely
red optical-near-infrared colours, e.g.\ $(I-K)\geq 4.0$, and as such
are termed Extremely Red Objects (EROs).  This strong spectral decline
could either result from a break in the SED or the suppression of bluer
light due to strong dust absorption.  As we look at more distant
galaxies, the first spectral break strong enough to produce such a red
colour is the 4000\AA\ break, which falls between the $I$ and $K$
passbands at $z>1$.  Progressively bluer features, such as the Balmer
break (3625~\AA) and Lyman-limit (912~\AA), will fall in the relevant
wavelength range for higher redshift sources, although these will
typically be fainter than those systems selected on the basis of the
4000\AA\ break.  Similarly, the dusty systems within the ERO population
are also expected to lie at high redshifts, due to the strong
wavelength dependence of dust reddening combined with the redshifting
of UV light into the optical passbands.  ERO surveys thus enable us to
identify both the most and least active galaxies at $z\gsim 1$, and
hence provide a powerful probe of obscured star formation at $z\gsim 1$
\citep[][hereafter S02]{Smail:2002dv} as well as the nature of the
evolved descendants of some of the earliest phases of star formation in
the Universe \citep{1996Natur.381..581D,1997Natur.390..377Z}.

Unfortunately, the mixed nature of the ERO population, compounded by
the strong angular clustering of one or both sub-populations
\citep{2000A&A...361..535D}, has led to a vigorous debate about the
relative importance of the evolved and active components in the overall
ERO population (\citealt{2002A&A...381L..68C}; S02).

The deepest ERO surveys undertaken show a rapidly increasing surface
density of sources as a function of magnitude limit down to $K\sim 19$.
Fainter than this, the increase in the numbers of EROs with apparent
magnitude slows dramatically
\citep{2002MNRAS.330....1S,2002MNRAS.332..617F}.  This transition is
very abrupt and suggests that there may be a profound change in the
nature of the EROs fainter than $K\sim 19$, perhaps associated with a
variation in the relative proportion of passive and active systems at
fainter magnitudes (\citealt{2002MNRAS.330....1S}; S02).  To
investigate this possibility we need an observational test to
distinguish between the different sub-classes of EROs.  There are three
approaches which have been used to attempt to differentiate passive
from dusty/active EROs: photometric tests, dust-insensitive star
formation indicators and morphological classification.

The photometric classification of EROs was popularised by
\citet{2000MNRAS.317L..17P}, using $J$-band photometry to separate the
two sub-classes on the $(I-K)$--$(J-K)$ plane.  Subsequent application
of this test to various samples has led to the conclusion that the
population at $K\lsim 20$ is roughly equally split between the
dusty/active and passive sub-classes
(\citealt{2002MNRAS.329L..57M,2002A&A...396..847V,2002MNRAS.330....1S};
see also \citealt{2002A&A...381L..68C}).  Unfortunately, without an
independent test of this classification scheme it is hard to know how
reliable these conclusions are.

Searches for millimetric and radio emission from samples of EROs have
also been used to attempt to estimate the proportion of dust-obscured
star-forming EROs in the overall population
(\citealt{2002A&A...383..440M}; \citealt{2002ApJ...577L..83W}; S02).
However, these star-formation indicators, while being insensitive to
dust obscuration in the galaxies, are only capable of detecting the
most vigorously star-forming systems at $z\gsim 1$.  As a result they
provide only a lower-limit of $>10$\% \citep{2002A&A...383..440M} and
$>30$\% (S02) on the proportion of star-forming systems in the ERO
population.  S02 also compare the colours of their radio-selected
star-forming EROs with those expected for dusty star-forming and
passive galaxies and find that most fall on or within the model
classification boundary for dusty galaxies, providing some support for
the \citet{2000MNRAS.317L..17P} scheme.

The final observable which has been used to differentiate between
evolved and active ERO populations is their morphologies.  It is
commonly assumed that the passive EROs will have elliptical or at least
early-type \citep{2000A&A...364...26M,2002MNRAS.333L..16S}
morphologies, while star-forming galaxies are expected to exhibit
either a disk-like, disturbed or obviously interacting morphology
\citep{1999ApJ...519..610D}.  Morphological studies of very red
galaxies have indeed shown that some have elliptical morphologies
\citep{1999MNRAS.309..208M,1999ApJ...524L..27T}.  However, results on
statistical samples of EROs have produced more mixed conclusions, with
\citet{2000A&A...364...26M} and \citet{2002MNRAS.337.1282R} claiming a
high early-type fraction and more recent work by
\citet{2002MNRAS.330....1S} and \citet[][hereafter
YT03]{2003ApJ...586..765Y} arguing for a more equal mix of early-type
and later-type/disturbed systems.  By comparing this classification
scheme with those described above it is possible to test the
reliability of the different approaches for a large sample of EROs
\citep[c.f.][]{2002MNRAS.330....1S}.

We highlight the study of YT03 which has very similar goals to those
described in this paper.  Their work utilised high resolution {\it HST}
WFPC2 F814W-band imaging to morphologically classify a sample of 115
$(I_{814}-K)\geq 4.0$ EROs selected from ground-based $K_s$-band
imaging to a 5-$\sigma$ depth of $K_s \lsim 18.7$. The authors visually
classified their objects and found a mix of around 30:64:6 for bulge
dominated, disk dominated and unclassified classes, with around 17\% of
the sample showing signs of merger/strong interaction.  They showed
their visual estimates to be in generally good agreement with an
automated bulge$+$disk decomposition technique.  The key improvement of
our study over that of \yt\ is that our near-infrared imaging is
sufficiently deep to probe beyond the break in the ERO counts at $K\sim
19$. We also improve upon their work through the addition of
multicolour photometry to test the different techniques for
subclassification of the ERO population.

We discuss our near-infrared observations of a selected sample of deep
{\it Hubble Space Telescope} WFPC2 images in \S2, present our analysis
of these in \S3 and the results which this provides are presented and
discussed in \S4, before giving our conclusions in \S5.  Throughout we
assume a cosmology with $\Omega_m=0.3$, $\Omega_\Lambda=0.7$ and
$H_o=70$\,km\,s$^{-1}$\,Mpc$^{-1}$.

\section{Observations and data reduction}
\label{sec:observ-data-reduct}

A common misconception is that the key to a successful ERO survey is to
obtain deep NIR imaging -- in fact, the observationally most demanding
aspect is achieving the necessary depth in the optical to identify that
a galaxy is an ERO. For this reason, we have chosen to concentrate our
survey on fields for which deep, high-quality optical imaging already
exists. These fields come from the Medium Deep Survey
\citep{1994ApJ...437...67G} which consists of over 500 deep {\it
  HST}/WFPC2 images of intermediate/high-Galactic latitude blank
fields. We selected a subsample of $\sim100$ fields from the parallel
and pointed samples from the survey.  These were selected by requiring
only that the total exposure time in the F814W filter be greater than
4.0ks, that the primary target of any parallel observations was not a
galaxy cluster or that, if the observations were pointed, that they
were not targeting an extragalactic source (to ensure representative
extragalactic regions) and that they be suitably placed for northern
hemisphere follow up (Dec. $>-10^\circ$). These then represent
high-resolution (0.1 arcsec FWHM) and very deep ($I_{lim}\sim25$--26)
images of random areas of the extragalactic sky.

\subsection{Near infrared observations}

Near infrared imaging was obtained using the Isaac Newton Group Red
Imaging Device \citep[INGRID,][]{ingrid} on the 4.2m William Herschel
Telescope (WHT)\footnote{Based on observations made with the William
  Herschel Telescope operated on the island of La Palma by the Isaac
  Newton Group in the Spanish Observatorio del Roque de los Muchachos
  of the Instituto de Astrofisica Canarias.}.  INGRID is a 1024$^2$
HAWAII-2 array at the bent Cassegrain focus of the WHT giving a 0.238
arcsec pixel$^{-1}$ scale and a 4.1$\times$4.1 arcmin field.

Observations were made over 12 nights, 2000 November 13--15, 2000
December 9--11 and 2001 May 1--6.  One night was lost due to the
instrument being unavailable. A total of 55 fields were observed in the
\ks-band and 30 of these in the $J$-band.  Four of the nights were
non-photometric, and we are in the process of obtaining calibration
data for these.  For the rest of this paper, we consider the bulk of
the \ks-band data, 35 fields covering 206 arcmin$^2$ ({\it HST} $+$
$K$-band coincident imaging) which are well calibrated.  A summary of
these fields is given in Table~\ref{table:fields}.

Each MDS field was imaged using a 9 point dither pattern with
exposures of around 60s in \ks\ and \j~exposures of
around 120s. Each \ks\ image was made up of $4 \times 15$s exposures
coadded in hardware. The 9 point dither pattern itself was moved
around the target field, in order to avoid bright objects
falling on the same pixels.  Total integration times are typically
2.8ks and are listed in Table~\ref{table:fields}.

\begin{table*}
\caption{Table of observations. $^a$ -- Field names are derived from
  MDS catalogue names. Fields suffixed XA and XB
  represent two overlapping {\it HST} exposures, with one corresponding
  $K$-band exposure targeted at the midpoint of the two; $^b$ -- Target name is target of
  {\it HST} observation for pointed fields; PAR indicates a pure parallel field.}
\label{table:fields}
\begin{tabular}{lcccccccl}
\hline
Field$^a$ & R.A. & Dec & F814W  & F814W & K & K & K & Target$^b$ \\
ID & \multispan2{(J2000)} & T$_{exp}$ (ks) & 5-$\sigma$ limit & T$_{exp}$ (ks) &
5-$\sigma$ limit & seeing (arcsec) & \\
\hline
UFG00   & 00~18~29.8 & +16~20~39.4 &   4.7 & 25.06 &   2.8 & 20.09 &  0.73  & PAR\\
UHG00   & 00~20~14.5 & +28~35~22.6 &   5.6 & 25.75 &   2.6 & 20.29 &  0.94  & PAR\\
UEH02   & 00~53~33.7 & +12~50~24.1 &   4.2 & 25.58 &   2.8 & 20.11 &  0.95  & PAR\\
UJH01   & 01~09~05.5 & +35~35~35.9 &   4.2 & 25.27 &   2.8 & 20.11 &  0.73  & PAR\\
UBI04   & 01~10~03.3 & $-$02~25~19.8 &   5.5 & 25.64 &  2.8 & 20.25 &  1.10  & PAR\\
UFJ00   & 02~07~03.3 & +15~26~00.3 &   4.2 & 25.51 &   2.1 & 20.14 &  0.86  & PAR\\
UGK00   & 02~38~49.5 & +16~45~25.1 &   5.4 & 25.36 &   2.8 & 20.34 &  0.94  & PAR\\
U2IY2   & 03~02~37.2 & +00~12~32.4 &   6.4 & 25.49 &   2.8 & 20.21 &  0.76  & FIELD-030233+00125\\
U2V12   & 03~02~42.6 & +00~06~27.9 &   6.7 & 25.85 &   2.8 & 20.31 &  0.91  & FIELD-030239+00065\\
U2IY1   & 03~02~46.2 & +00~13~04.1 &   6.4 & 25.64 &   2.8 & 20.15 &  0.86  & FIELD-030243+00132\\
U2V19   & 03~38~35.4 & $-$00~12~38.3 &   6.7 & 25.77 & 2.8 & 20.28 &  0.97  & FIELD-034112-00035\\
U2V17   & 03~40~58.1 & +00~03~57.4 &   6.5 & 25.87 &   2.8 & 20.19 &  1.04  & FIELD-034101+00030\\
U2V18   & 03~41~12.2 & +00~00~28.4 &   6.7 & 25.91 &   2.4 & 21.10 &  0.68  & FIELD-034115+00002\\
UPJ00   & 06~52~46.3 & +74~20~43.6 &   4.2 & 25.37 &   2.8 & 21.06 &  0.69  & PAR\\
UQJ10   & 07~27~36.9 & +69~04~56.3 &   4.1 & 25.29 &   2.8 & 20.09 &  0.69  & PAR\\
UQK02   & 07~41~33.5 & +65~05~28.7 &   6.6 & 26.05 &   2.1 & 20.43 &  0.83  & PAR\\
UQL00   & 07~42~41.4 & +49~43~21.3 &   4.2 & 25.20 &   2.8 & 20.23 &  0.73  & PAR\\
UOP00   & 07~50~47.1 & +14~39~48.9 &   4.2 & 25.38 &   2.8 & 20.46 &  1.03  & PAR\\
USP00   & 08~54~15.6 & +20~02~47.0 &   4.2 & 25.28 &   2.0 & 19.85 &  0.66  & PAR\\
UPS00   & 09~09~58.5 & $-$09~26~51.9 &   5.6 & 25.76 &  2.8 & 20.17 &  1.20  & PAR\\
UVM01   & 09~39~34.7 & +41~33~37.4 &   4.6 & 25.23 &   2.8 & 20.03 &  0.76  & PAR\\
UWP00   & 10~02~24.8 & +28~51~01.8 &   8.4 & 26.13 &   2.8 & 21.96 &  0.81  & PAR\\
UUS00   & 10~04~52.9 & +05~15~52.3 &   4.6 & 25.45 &   2.8 & 20.53 &  0.71  & PAR\\
UST00   & 10~05~47.9 & $-$07~40~39.9 &  23.1 & 26.24 &  2.8 & 20.08 &  1.13  & PAR\\
U2RJ1XB & 11~48~47.8 & +10~55~50.3 &   5.3 & 25.19 &   1.9 & 20.17 &  0.74  & PAR\\
U2RJ1XA & 11~48~50.7 & +10~56~36.6 &   6.9 & 25.66 &   1.9 & 20.17 &  0.74  & PAR\\
U2H92   & 13~12~14.6 & +42~45~30.6 &  15.6 & 26.67 &   1.6 & 20.68 &  0.87  & SSA13\\
UY401   & 14~35~30.7 & +25~17~30.8 &   8.0 & 26.29 &   2.8 & 21.34 &  0.75  & PAR\\
U2AY2   & 15~58~49.7 & +42~06~18.3 &  25.2 & 26.70 &   2.2 & 20.44 &  0.88  &  DEEP-SURVEY-FIELD\\
UMD0EXB & 21~50~32.7 & +28~49~51.8 &   8.4 & 25.95 &   2.8 & 20.74 &  0.63  & PAR\\
UMD0EXA & 21~50~33.9 & +28~48~29.0 &   5.6 & 25.67 &   2.8 & 20.74 &  0.63  & PAR\\
UMD07XB & 21~51~04.1 & +29~00~33.1 &   9.6 & 25.84 &   2.3 & 21.04 &  0.88  & PAR\\
UMD07XA & 21~51~09.8 & +29~00~39.6 &   8.7 & 26.35 &   2.3 & 21.04 &  0.88  & PAR\\
UMD0D   & 21~51~25.6 & +28~43~49.2 &   5.6 & 25.43 &   2.8 & 20.44 &  0.80  & PAR\\
U2H91   & 22~17~35.7 & +00~14~07.7 &  28.8 & 26.37 &   2.4 & 20.22 &  0.83  & SSA22\\
U2V16   & 22~17~37.8 & +00~17~14.3 &   6.7 & 25.84 &   3.4 & 20.06 &  0.95  & FIELD-221736+00182\\
U2V14   & 22~17~59.3 & +00~17~15.5 &   6.7 & 25.59 &   2.8 & 20.07 &  0.78  & FIELD-221755+00171\\
UED01   & 23~19~52.2 & +08~05~31.6 &   5.6 & 25.48 &   2.8 & 20.20 &  0.92  & PAR\\
\hline
\multispan3{\hfil Median values } & 6.4  & 25.64 & 2.8 & 20.25 & 0.83 & \\
\hline
\end{tabular}
\end{table*}

The INGRID data were reduced using a custom written pipeline, available
from {\tt http://star-www.dur.ac.uk/$^\sim$dgg/ipipe/}. The pipeline
uses standard {\sc iraf}\footnote{{\sc iraf} is distributed by the National Optical
  Astronomy Observatory which is operated by AURA Inc.\ under contract
  with the NSF.} and {\sc starlink} routines, and the reduction procedure is
as follows.

\subsubsection{Initial processing}
 
The first step in the processing was to construct a bad pixel mask and
apply a correction for an offset in the exposure time present in early
versions of the software controller\footnote{ {\tt
    http://www.ing.iac.es/Astronomy/instruments/ingrid/ ingrid\_timing.html}}.
Dark subtraction was found to be unnecessary for most frames, but a
master dark frame was subtracted if residual structure was seen.  The
data were flatfielded and sky subtracted using the {\it in-field
  chopping}, or {\it moving flatfield} technique of \citet{cowie90},
using a running median of 8 temporally adjacent frames.

\subsubsection{Mosaicking}
Image registration and mosaicking was performed entirely using the
{\sc starlink} software {\sc ccdpack}. The first image of each target was
taken as the reference and the relative offsets of the others
calculated. This was achieved using the tasks {\sc findobj}, 
{\sc findoff} and
{\sc register} to find objects in common between frames and calculate the
offsets to sub-pixel accuracy.  The images were then geometrically
transformed with subpixel shifts and bilinear interpolation to conserve
flux using {\sc tranndf}, and finally combined using a 3-$\sigma$ clipped mean
within the {\sc makemos} task.

The mosaic so produced is regarded as a first pass mosaic.  In order to
improve sky subtraction, a mask is made for all the astronomical
objects in this image.  \sex\ v2.2.2 \citep{sex} was used to detect and
mask objects by making a CHECKIMAGE with the OBJECTS option.

The sky subtraction step is then repeated by first applying the
de-registered mask to the images going into the local flatfield. In
this way, objects too faint to be found in the individual exposures
which would otherwise bias the estimate of the sky level and lead to
oversubtraction of the sky can be successfully rejected (M.\ Currie,
priv.\ comm.).  The mosaicking step is then repeated with these second
pass sky subtracted images to make the final mosaic.

In addition to the image mosaic, an exposure map is constructed by
summing the number of images going into each pixel.  This is used to
deal with the uneven noise properties of the mosaics.  An image with
uniform pixel to pixel noise is generated by multiplying the mosaic by
the square root of the exposure map. However, for the current dataset,
the field of view of INGRID is very well matched to the WFPC2, allowing
the latter to be entirely contained within the area of the former for
any roll angle.  Thus, no region of the WFPC2 field sees less than
100\% of the near-infrared (NIR) exposure time, except the three pairs
of fields suffixed `XA' and `XB'.  These are fields where two
overlapping MDS pointings exist, and the INGRID pointing lies midway
between.  Only NIR data with more than 50\% of the total exposure time
is considered in these three fields.

\subsubsection{Non-photometric data correction}
\label{sec:non-photometric-data}
In the discussion of the mosaicking process above, no correction has
been made to the photometric zeropoints for each individual frame going
into the mosaic.  This is because the extinction in the NIR is low
($\lsim 0.08$ magnitudes/airmass) and the fields were typically observed for
around 2.8ks at low airmass, therefore the change due to the
extinction variation from image to image is negligible. This is not
the case for non-photometric data.  However, it is straightforward to
correct all the exposures of one field to the same zeropoint, as the
interval between exposures is short, and the dither pattern is small,
so there are many objects in common between frames.  

To correct for non-photometric data, the first frame of a field was
again taken as the reference image.  The first pass mosaic was made as
before, but when \sex\ was run on it to make the object mask, a
catalogue of the 20 brightest objects in the field was output. After
the second pass sky subtraction, aperture photometry was performed on
these 20 objects and the optimal photometric zeropoint shift between
each exposure and the reference calculated using the {\sc linfit} task
in {\sc iraf}. Typical shifts were $\sim 0.1$ mag. The validity of this
approach is verified through comparison with repeat observations made
in photometric conditions, and with external photometry, described
next.

\subsubsection{Photometric and astrometric calibration}
Calibration on to the UKIRT $K$-band system was performed using UKIRT
Faint Standard (FS) stars \citep{ukirtfs}.  Several such stars were
observed at the start and end of each night. The accuracy of the
calibration was checked against the 2MASS All-Sky Point Source
Catalogue\footnote{This publication makes use of data products from the
  Two Micron All Sky Survey, which is a joint project of the University
  of Massachusetts and the Infrared Processing and Analysis
  Center/California Institute of Technology, funded by the National
  Aeronautics and Space Administration and the National Science
  Foundation.}.  Around 60\% (including non-photometric nights) of the
fields possessed sufficient bright objects in common with 2MASS to
obtain an accurate test of the calibration.  The minimum number of
objects used was five, and more typically 10 - 20.  Where the data
overlapped, the agreement was found to be typically better than 0.05
mags. This also showed that zero point on non-photometric nights did
not differ from the nominal zero point estimated from the standard
stars by more than $\sim 0.3$ mags. We stress that we do not use data
where the absolute calibration is more uncertain than $\sim$0.1 mags.
All passbands are corrected for Galactic reddening using the maps of
\citet{1998ApJ...500..525S}.  Astrometry was performed against the
USNO-A2.0 astrometric catalogue \citep{usno} using the package {\sc
  wcstools} \citep{wcstools} to automatically calculate the World
Coordinate System (WCS) of the image, and found to be accurate to
better than 1 arcsec.

\subsection{{\it HST} imaging}
\label{sec:it-hst-imaging}
The {\it HST} images were retrieved from the ST-ECF Association
archive. The archive takes all available associated data defined
by the pointing of the telescope, performs an on-the-fly recalibration
(OTFR) and co-adds these chip by chip.  The only additional processing
necessary was to combine the four WFPC2 chips into a single mosaic. In
so doing, the smaller pixel scale of the PC chip is resampled to that
of the three WF chips.  This was done using the {\sc iraf} {\sc stsdas}
task {\sc wmosaic}, which also corrects for the geometric distortion
within the instrument.  Photometric zeropoints were taken from the
headers produced by the OTFR and are calibrated to the Vega system as
described in \citet{1995PASP..107.1065H}.  The F814W Vega system is
very close to the Cousins $I$-band and shall be referred to as \i814
~henceforth.

\subsection{Image alignment and photometric catalogue}
\label{sec:phot-catal}
The seeing in the ground-based data was automatically determined by
profile fitting to stellar objects in each frame. For each field, the
{\it HST} data were then matched using Gaussian convolution to the
seeing of the passband with the poorest seeing. All passbands were
aligned to the \ks-band image and resampled to the INGRID pixel
scale\footnote{Hereafter, {\it pixel} shall refer to this common pixel
  scale of 0.238 arcsec unless explicitly stated otherwise.}, again
using the {\sc starlink} package {\sc ccdpack} to perform automated
registration.  The field of INGRID is astrometrically very flat, as is
the WFPC2 field after geometric correction with {\sc wmosaic}; thus the
residual offsets between objects in the different passbands were
typically at the level of one pixel.  Although the internal astrometry
of the WFPC2 device is very good, the external accuracy is somewhat
lower, thus the WCS that we adopt in the catalogues is that of the
\ks-band image.

Photometry is performed on coaligned images after convolution to the
same PSF.  These shall be referred to as the {\it convolved} images.
Also, in order to exploit the exquisite resolution of the {\it HST}, we
will wish to use the images prior to this transformation. We shall
refer to these images as the {\it non-resampled} images. The {\sc
  ccdpack} software allows the mapping between our different coordinate
systems to be readily stored, and a provides convenient way to
transform between them.

\sex\ was used to detect objects brighter than the $\mu_K = 21.2$ mag.\ 
arcsec$^{-2}$ isophote, with at least 4 connected pixels (0.23
arcsec$^2$), after filtering with a Gaussian with a FWHM of 4 pixels
(0.95 arcsec).  This limit is determined from our shallowest fields, in
order to ensure high uniformity in the object catalogues.  Object
detection was run on the uniform noise mosaic, described above.
Photometry is performed by running \sex\ in double image mode, using
the detection image to define the location of photometric apertures and
measuring magnitudes from each convolved passband in turn (note that
$K$-band magnitudes were measured from the \ks-band image, {\it not}
from the \ks-band uniform noise image).  An aperture diameter of 2.5
times the FWHM of the seeing was used.  This ensures high signal to
noise within the aperture \citep{lcg91}, and is small enough (typically
around 2 arcsec) to avoid contamination from neighbouring objects, due
to the good seeing of our data. We verified that our choice of
photometric aperture is robust, by also examining colours defined by a
fixed 2 arcsec aperture, a 4 arcsec aperture, and `total' colours
measured using the \sex\ BEST\_MAG in each band (where the BEST\_MAG
aperture is defined from the \ks image).  It was found that all
these measures are consistent, but that our adopted approach typically
yields the colour with the smallest error.

Total magnitudes are measured for the $K$-band using the \sex\ 
BEST\_MAG.  5-$\sigma$ limiting magnitudes within these apertures were
calculated using \sex's estimate of the signal to noise for each object
detected (Table~4).  The adopted 5-$\sigma$ limiting aperture magnitude
for the survey is $K\leq20.0$, and only objects brighter than this are
considered.

To assess completeness, a set of simulations was run first by inserting
artificial point sources and then artificial elliptical and disk
galaxies (of the typical size of faint galaxies in the survey) into our
$K_s$-band images. Our detection procedure was then repeated on these.
The typical completeness for point sources is greater than 80\% at
$K=20.0$, and a similar completeness level is reached around 0.3 mag
brighter for objects with de Vaucouleurs profiles.  The simulations
show we are 80\% complete for compact sources at K$=$20.0; while they
also show that the sample is 100\% complete for point sources around
K$=$19.4, de Vaucouleurs profiles at K$=$19.1 and the intermediate case
of exponential disk profiles is 100\% complete around K$=$19.2.  Since
our primary aim is to interpret the morphologies of EROs in the context
of our survey and previous imaging surveys, which are typically
selected to a limiting 5-$\sigma$ depth, we apply this selection to our
sample.  A discussion of the completeness for different morphological
types will be considered in future work, when comparing our data with
galaxy formation models.  In order to assess the spurious detection
rate, the detection images were multiplied by $-1$ and the detection
and measurement procedure repeated.  No false detections brighter than
$K=20.0$ were found.

\section{Analysis}

Our analysis will proceed by using our K-selected galaxy survey to
isolate Extremely Red Objects based on their $(I_{814} - K)$ colours.
We will use the superlative resolution of the {\it HST} imaging to
morphologically classify these EROs.  Both qualitative visual
classification and automated machine-based classification schemes will
be used. We shall then compare our photometry, number counts and
morphological fractions with previous work.

\subsection{ERO sample selection}

EROs were identified by selecting all objects matching the common
definition (e.g. S02, \citealt{2002ApJ...577L..83W},
\citealt{2003astro.ph..3206R}, \yt) of an ERO of $(I_{814}-K) \ge 4.0$.
Our goal is to determine the morphological mix of such EROs to
K$\sim$20. The astrophysical motivation behind this colour selection is
described in detail in \citet{2000MNRAS.317L..17P} and references
therein.  Briefly, this colour cut corresponds to the expected colours
for a z$\sim$1 passively-evolving, old stellar population, but is much
redder than the expected colours of normal field galaxies; only
high-redshift, heavily dust-reddened, massive starbursts can have
similar colours.  Therefore, this colour cut isolates the most and
least actively star-forming galaxies at z$\gsim$1.  

\begin{figure*}
\centering
\includegraphics[width=85mm,clip=t]{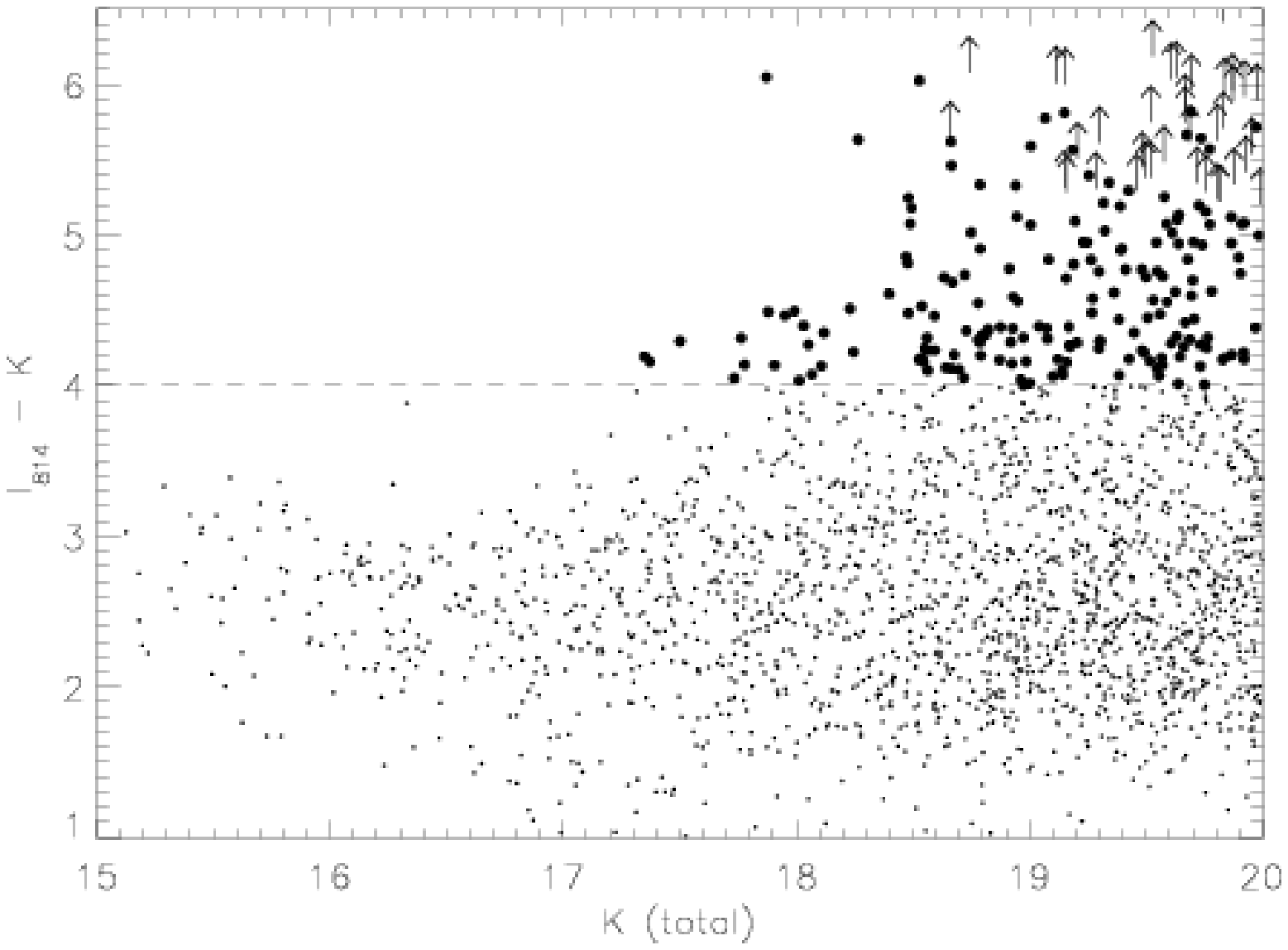}
\includegraphics[width=85mm,clip=t]{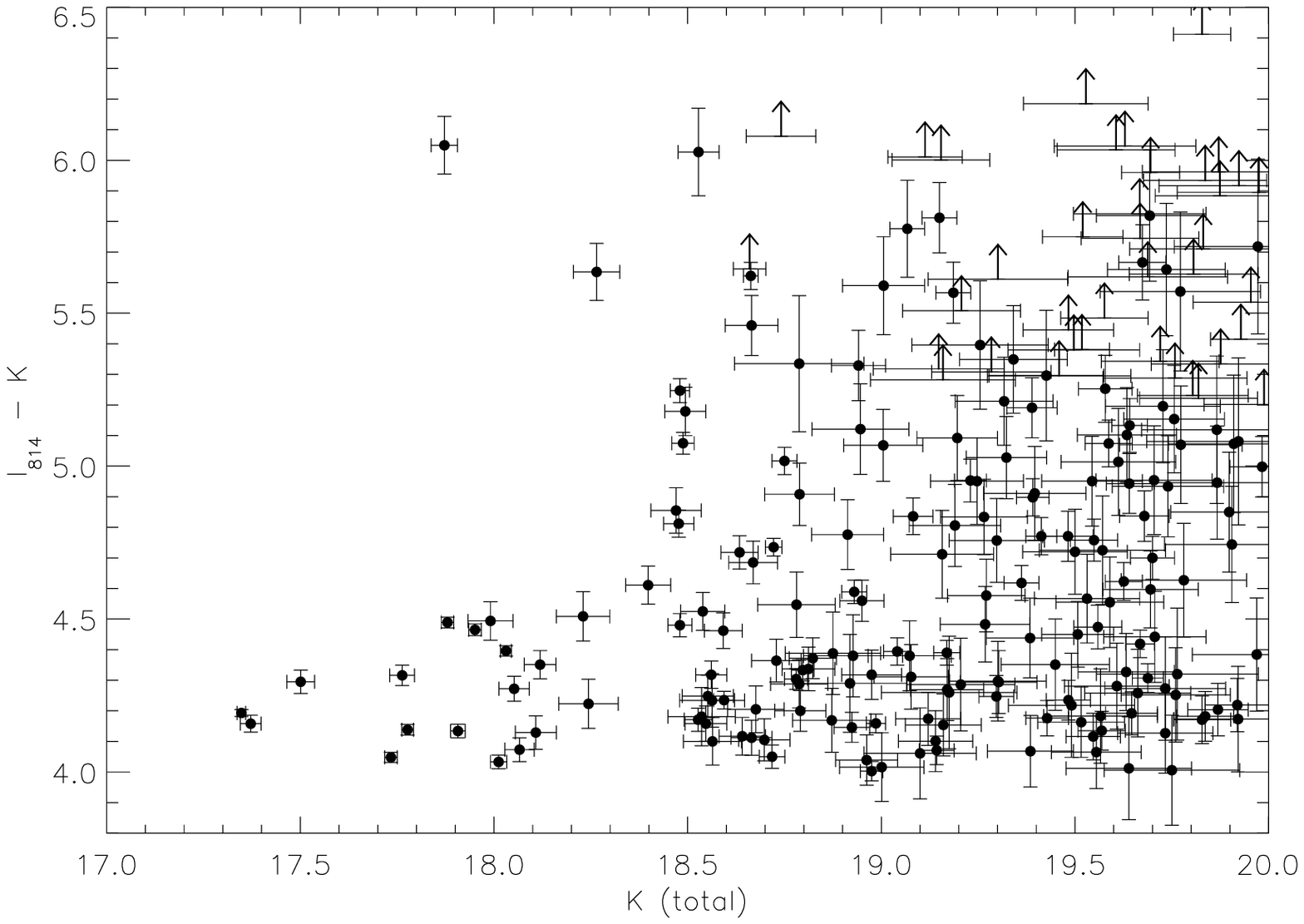}
\caption{Colour-Magnitude diagrams for all galaxies in the survey
  (left panel) and for EROs (right panel) defined by  $(I_{814}-K)\geq 4.0$.
Error bars are omitted from the left plot for clarity and the broken
line illustrates the $(I_{814}-K)\geq4.0$ ERO selection criterion. EROs are
highlighted with bold points.  Note the rapid increase in the
proportion of the galaxy population in the ERO class at $K>18$ (4\%), 
followed by a more constant distribution fainter than $K\sim 19$ where
the EROs contribute  12\% of the total population. 
}
\label{fig:cmd}
\end{figure*}

Fig.~\ref{fig:cmd} shows the colour-magnitude diagram for the whole
$K$-selected survey.  Our ERO catalogue was visually inspected and a
small number of sources were rejected as being false detections, either
being saturated stars or objects lying in the periphery of bright
stars.  No other false detections were found.  We emphasize that for
our entire survey, the NIR data used is highly uniform across the
entire WFPC2 field of view and does not suffer from reduced signal to
noise at the edges, as can be the case for surveys undertaken with
smaller NIR imagers (e.g.\ \yt). We class objects with a \sex\ 
CLASS\_STAR~$\le 0.97$ as galaxies.  Only one of our EROs would be
classed as stellar by this definition (ERO082). We include this object
for completeness (it may still be extragalactic, e.g.\ AGN, and will be
morphologically classified from WFPC2 imaging later) but note that its
inclusion has a negligible effect on our results.

We identify 224 sources to K$\leq20.0$ with $(I-K)\geq 4.0$ and 83 with
$(I-K)\geq 5.0$ across the 206 arcmin$^2$ of our survey.

\subsection{Morphological analysis}

Using the exquisite resolution of the WFPC2 we can morphologically
classify the EROs in our survey.  This classification will proceed in
two ways: we will use a scheme based on visual inspection of the images
and also a quantitative system using machine-based measurements of the
central concentration of each object.  Each method has its own
advantages and disadvantages: visual classification is somewhat
subjective, but is better suited to deal with unusual objects, not
well-fit by a model; whereas automated classification should be
reproducible for a given set of input parameters, but unusual objects
may lead to catastrophic misclassifications -- e.g.\ for a merging
system, the software might only fit one subcomponent, considering it to
be an isolated system.

For the morphological analyses, the non-resampled F814W images were
taken,and residual cosmic rays were rejected using the {\sc iraf} task
{\sc craverage}. Next the images were rebinned 2$\times$2 to increase
the per pixel surface brightness sensitivity.  The reduction in angular
resolution from this procedure does not adversely affect the
morphological classifications, but aids in classifying the lowest
surface brightness systems, which would otherwise be unclassifiable.

\subsubsection{Visual classification}

All the EROs were visually classified by one of us (IRS) following the
scheme in Table~\ref{table:visclass}. This scheme was devised to isolate
the broad classes of objects identified on a first-pass through 
the dataset.

This procedure used the same display software employed by the MORPHS
group when classifying faint galaxies in distant clusters
\citep{1997ApJS..110..213S}.  The software displays two images at
different stretches of a $10\arcsec\times10 $\arcsec~region around the
ERO from the non-resampled WFPC2 images, after they are smoothed with a
0.2\arcsec ~FWHM Gaussian to reduce the shot noise.  This was used in
conjunction with a hard-copy of each thumbnail image displaying the
\ks-band image of the ERO contoured over the F814W image to allow
isolated \i814 -band components within a single ERO to be easily
identified (e.g.  Fig.~\ref{fig:thumbs}).

\begin{table}
\caption{Visual classification scheme for the EROs, based on the WFPC2
  F814W imaging. Classes are motivated by the properties of the sample
  and we bin these to describe disk-like (2, 5, 6) and
  bulge-like (3 and 7) systems.}
\label{table:visclass}
\centering
\begin{tabular}{l|lcc}
\hline
Class & Description & \# & \% \\
\hline
\hline
0 & Blank or too faint to classify      &  43  &   19.20  \\
1 & Compact (small and peaked)  &  32  &   14.29  \\
2 & Compact, disk       &  36  &   16.07  \\
3 & Compact, symmetrical        &  26  &   11.61  \\
4 & LSB, disturbed      &   1  &    0.45  \\
5 & LSB         &   6  &    2.68  \\
6 & Obvious disk        &  42  &   18.75  \\
7 & Spheroidal  &   7  &    3.12  \\
8 & Merger      &  13  &    5.80  \\
9 & Amorphous   &  18  &    8.04  \\
\hline
\end{tabular}
\end{table}

We give the proportions of different morphological classes in our full
K$\leq$20 sample in Table~\ref{table:visclass}.  In a testament to the
depth of the {\it HST} imaging, less than 20\% of these very red and
very faint galaxies were unclassifiable in the WFPC2 images.  For the
remaining sources, 40\% are compact (although these can still exhibit
weak morphological features), a further 20\% are relatively
well-resolved, disk galaxies and the final 20\% comprise a mix of low
surface brightness (LSB), mergers, amorphous or clearly spheroidal
galaxies.  As we have stressed, these morphological classes were
motivated by the properties of the ERO sample and hence we will need to
combine and convert them to translate them into astrophysically
interesting classes.

\subsubsection{Automated classification}
\label{sec:autom-class}
Given the modest S/N of typical EROs in our WFPC2 imaging, we have
chosen in our quantitative morphological analysis to concentrate on two
simple measures of the light distribution within the EROs.  These are
the concentration index, $C$, and mean surface brightness, $\mu$
\citep[][]{1994ApJ...432...75A,1996MNRAS.279L..47A,1996ApJS..107....1A}.
Measurements of $C$ for the ERO sample were made using a modified
version of \sex\ to calculate $C = F_{0.3} / F_{1.0}$ where $F_x$ is
the integrated flux within an elliptical aperture which contains a
fraction $x$ of the total isophotal area, and x$=$0.3 is the standard
definition \citep{1994ApJ...432...75A}.  The error on this measurement
was estimated from simulations. Grids of artificial galaxy images were
constructed using the {\sc iraf} {\sc artdata} package and inserted
into random blank sky regions of the non-resampled F814W image, and the
scatter in the measurement of $C$ for the same input profiles recorded
as a function of apparent brightness of the galaxy.  The measured
variation in $C$ with apparent brightness of the galaxy (for a fixed
light profile) in these simulations also allows us to investigate the
evolution in the morphological mix of the EROs as a function of
magnitude in a quantitative manner.

\subsection{External comparisons}

\subsubsection{K-band galaxy counts}

\begin{figure}
\centering
\includegraphics[width=85mm,clip=t]{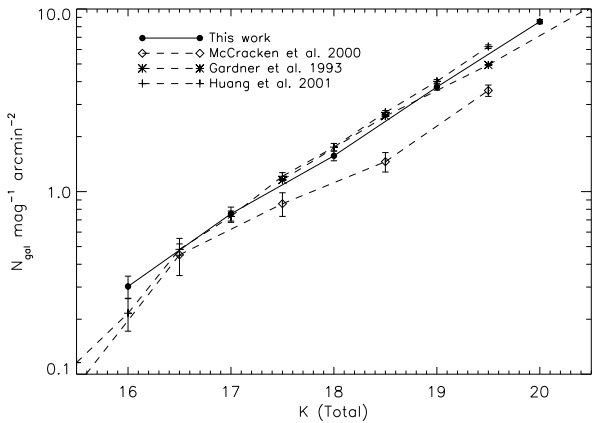}
\caption{Differential $K$-band galaxy counts in half-magnitude
  intervals, corrected for incompleteness. Error bars simply assume
  Poisson statistics and are thus likely to be underestimates of the
  true errors. Selected number counts from the literature are also
  plotted. Our counts are in good agreement with other surveys.  }
\label{fig:kcounts}
\end{figure}

As a check of our $K$-band photometry, in Fig.~\ref{fig:kcounts} we
compare our galaxy number counts (for all K-selected galaxies,
irrespective of colour) with a selection from the literature.  Our
counts are in good agreement, showing a steep rise with gradient
$\alpha=0.34\pm0.06$ to K$\sim$20, with an incompleteness-corrected
surface density of $(7.5 \pm 0.2)$ arcmin$^{-2}$ at K$=$20.

\subsubsection{ERO number counts}
\label{sec:ero-number-counts}
\begin{figure*}
\centering
\includegraphics[width=85mm,clip=t]{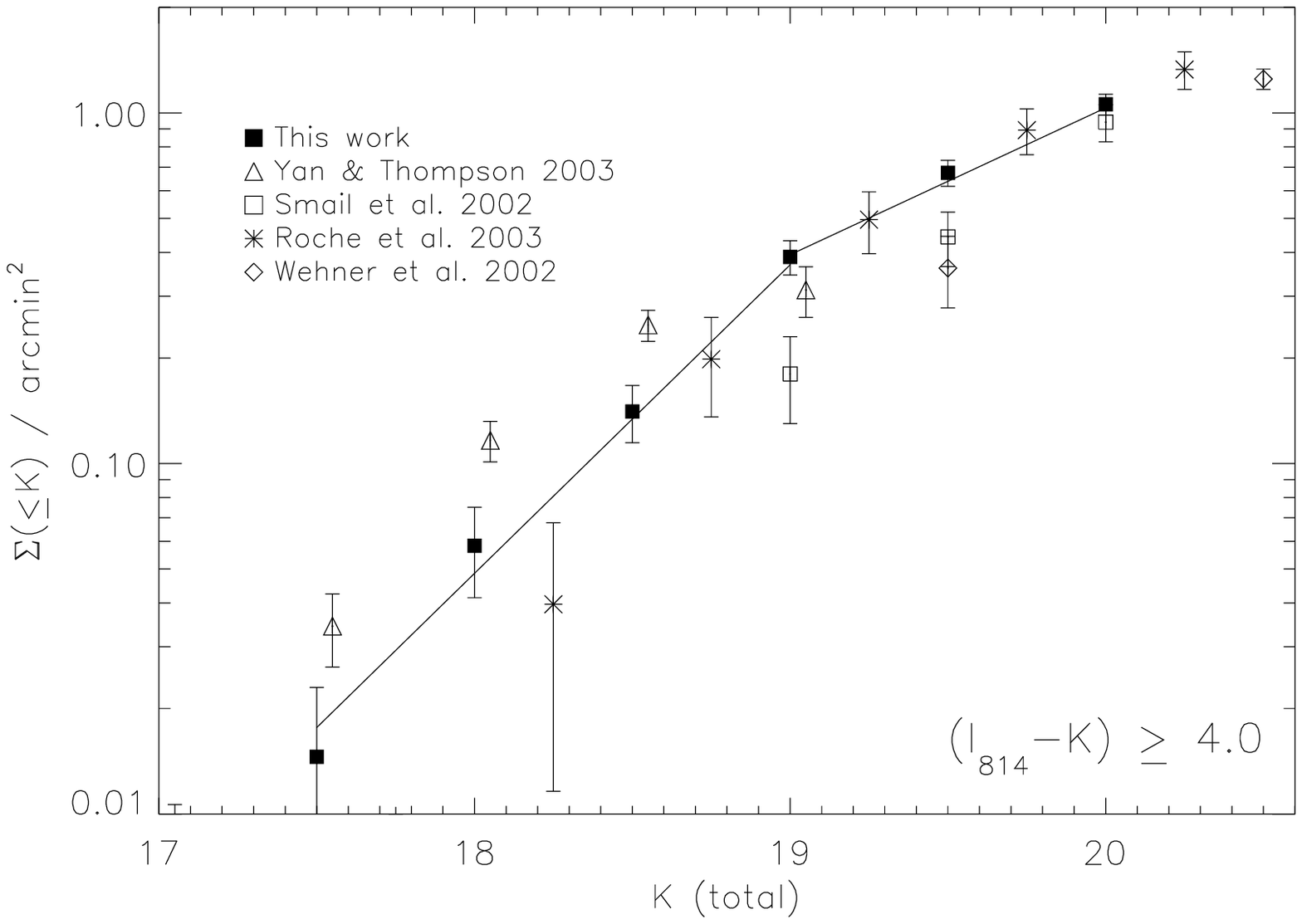}
\includegraphics[width=85mm,clip=t]{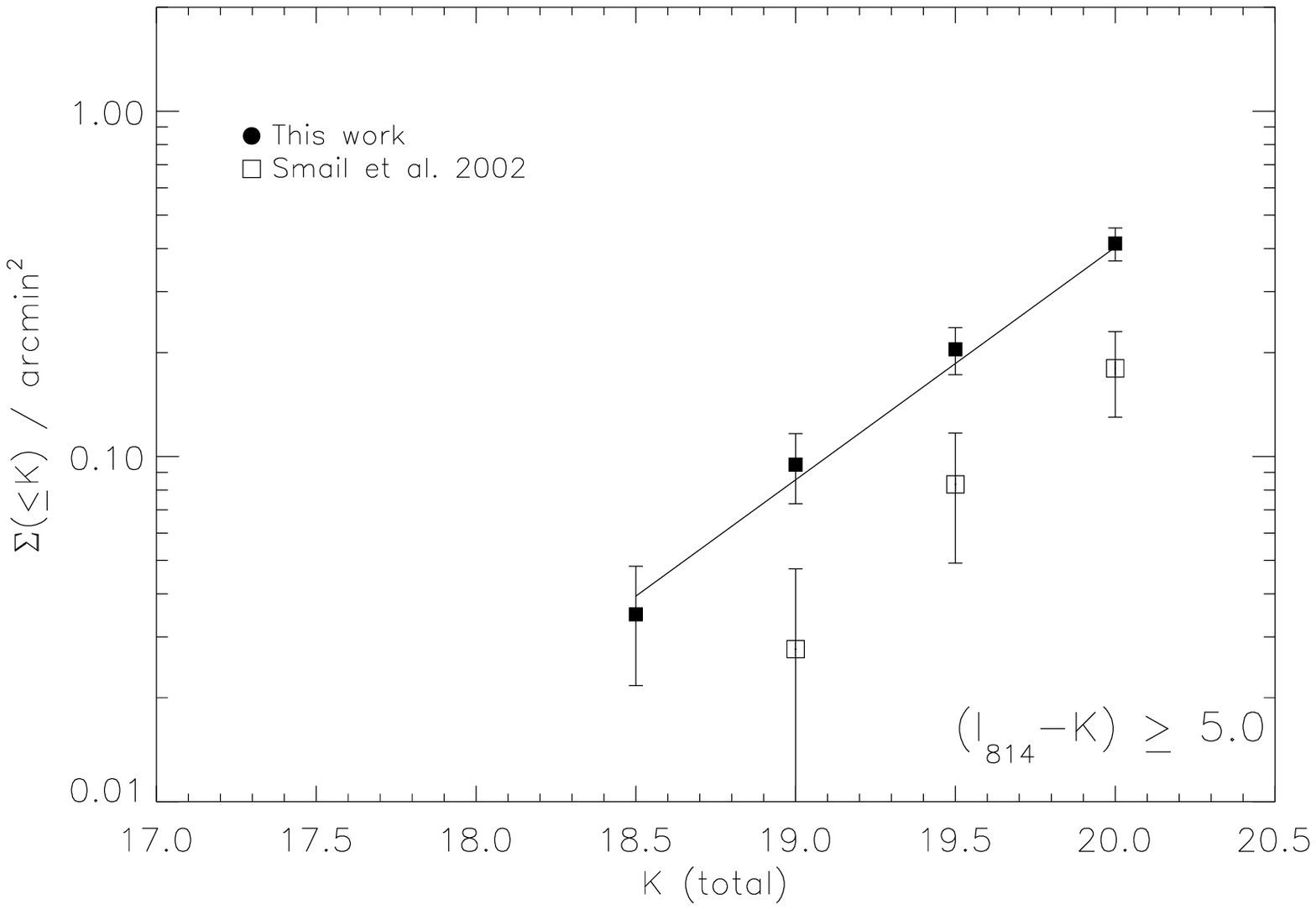}
\caption{Surface density of EROs. Left panel shows cumulative number
  counts for the full ERO sample selected with $(I_{814}-K)\geq4.0$.
  Overplotted are other ERO counts from the literature, with their
  selection criteria listed.  Right panel shows the redder subsample
  selected with $(I_{814}-K)\geq5.0$ compared with the redder subsample
  of S02 (priv. comm.) using the same selection criterion. Solid lines
  are our power law fits.}
\label{fig:counts}
\end{figure*}

Firstly we consider the number counts of $(I_{814} -K)\geq4.0$ selected
EROs in our survey.  We compare our number counts to other datasets
which use the same $(I_{814}-K)$ selection to remove any concerns about
differences in the samples due to the photometric selection (although
see \S\ref{sec:ero-sample-diff}).  This comparison is shown in the left
panel of Fig.~\ref{fig:counts}. Our cumulative ERO counts appear to show
a break around $K\sim 19$ and so we fit them using two power laws
($\log_{10}N(K_1<K<K_2) \propto \alpha K$), giving slopes of:
$\alpha=0.88\pm0.09$ for $K\leq19.0$ and $\alpha=0.42\pm0.19$ for
$19.0\leq K\leq20.0$. Also shown in the figure are the counts of
$(I_{814} -K)\geq4.0$ EROs from S02 (priv. comm.),
\citet{2002ApJ...577L..83W}, \citet{2003astro.ph..3206R} and \yt.

We see that our ERO counts are in excellent agreement with
\citet{2003astro.ph..3206R}, based on publicly available ESO imaging of
the CDFS/GOODS.  The S02 counts are marginally lower than the other
data, but note that the error bars are purely Poissonian errors and
likely to underestimate the true field-to-field variation
\citep{2000A&A...361..535D}.  At the faint end their number density
agrees well with both ours and those of \cite{2003astro.ph..3206R}. The
counts of \yt\ appear marginally higher but not significantly so.

We find an incompleteness-corrected surface density for $(I_{814}
-K)\geq4.0$ objects of ($1.14\pm0.08$) arcmin$^{-2}$ to $K\leq20.0$,
representing 17\% of the total galaxy population to this depth.  This
is in reasonable agreement with the value of ($0.94\pm0.11$) arcmin$^{-2}$
from S02 (using the same selection limits as ours, priv.  comm.).
\citet{2002ApJ...577L..83W} also find comparable numbers, but with
large field to field scatter between their three fields, each of which
uses a cluster lens to increase the sensitivity of their observations.
The data shown in the plot excludes one of their three cluster fields
(A2390) which seems particularly overdense in EROs.  This further
emphasizes the impact of field to field variations. The surface density
to the magnitude limit of the \yt\ survey also appears in good
agreement with our estimate, and is consistent with the S02 point.

Next we consider a redder subsample selected by $(I_{814}-K)\geq5.0$.
We find a surface density of such objects of $(0.41\pm0.05)$
arcmin$^{-2}$, or 5\% of the population to $K=20$.  The number counts
of $(I_{814}-K)\geq5.0$ objects are adequately described by a single
power law with a slope of $\alpha=0.70\pm0.02$.  Here we can only
compare with the $(I-K)\geq5.0$ counts of S02 (priv. comm.) who find a
surface density of $(0.18\pm0.05)$ arcmin$^{-2}$. Again we note that
the error estimates do not include a contribution from clustering. We
also note that this extreme colour cut is on the rapidly falling tail
of the colour distribution (encompassing $\sim$5\% of the population).
Slight photometric offsets between different surveys leading to
different colour measurements could cause a large number of objects to
be included or excluded.  For example, if our two surveys measure
colours systematically different by 0.2 mags (a reasonable calibration
offset between ground based surveys), this is enough to allow our
Poissonian error bars to overlap.

\subsubsection{Comparison with \yt\ photometry}

We compare our photometric measurements with those of \yt. We have 15
MDS fields in common, from our sample of 35 and their sample of 77 MDS
fields.  Firstly we take the ERO catalogue of \yt\ (their Table 2) and
search for objects which lie in our survey fields (Table
\ref{table:fields}).  We find 28 such objects (only 17 of which do we
classify as EROs).  Next we search our galaxy catalogues for entries
within 1\arcsec of the coordinates given by \yt. We match 26/28
objects.  The remaining two (their u2v16\#108 and u2v19\#30) lie
between the WFPC2 chips in our images and were thus masked from our
analysis. The ST-ECF associations archive only includes data taken at
very similar roll angles, so it is possible that \yt\ included
additional imaging not used in our analysis. Comparison of our F814W
exposure times for the fields in question (U2V16 and U2V19) show
identical exposure lengths to theirs.  Therefore we believe that their
inclusion of these objects in their catalogue is spurious.
 
\begin{figure}
\centering
\includegraphics[width=85mm,clip=t]{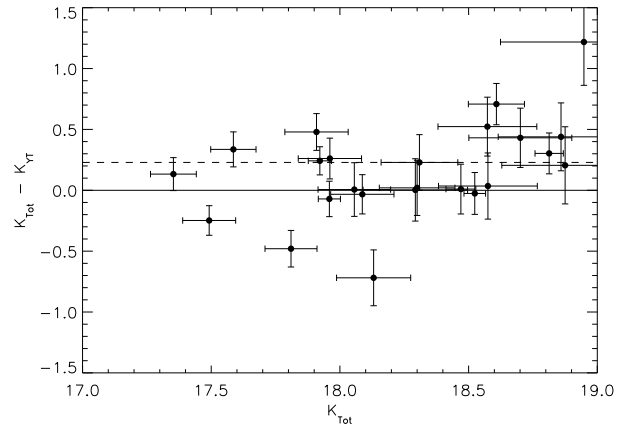}
\caption{The comparison between our $K$-band photometry and that of YT03 for
  galaxies in common.  We plot the difference between their large
  aperture magnitude and our total magnitude.  Even though our total
  magnitude should encompass more light than their aperture magnitude,
  there is evidence that their $K$-band magnitudes are systematically
  brighter, by around 0.3 mag (indicated by broken line). See text for
  discussion.}
\label{fig:ytphot}
\end{figure}

We compare our $K$-band total magnitudes with their large-aperture
(`total') magnitudes for these 26 galaxies and plot the results in
Fig.~\ref{fig:ytphot}.  For 10/26 (38\%) our estimates agree to
within 1-$\sigma$, however, for 13/26 (50\%) EROs \yt\ measure
considerably brighter $K$-band magnitudes, by 0.5--1 mag.  This offset
thus results in \yt\ measuring redder $(I_{814}-K)$ colours for these
galaxies and hence classifying them as EROs, whereas in our data they
are bluer than $(I_{814} - K) \geq 4.0$.  We visually examined the extreme
outliers and can find no obvious reason for the discrepancy.  For
example, one such outlier with $\Delta K = 1.21$ lies on the same WFPC2
chip as two other EROs which we measure to be the same brightness as
\yt~ to within 0.4 magnitudes. For these 26 galaxies, we measure a
median offset of $0.26\pm0.07$. This slight photometric offset would
bring their ERO number counts into closer agreement with ours and
\citet{2003astro.ph..3206R}.  We emphasize that this comparison only
uses data taken under photometric conditions and so cannot be affected
by corrections for non-photometric images.  We attribute the
disagreement to the quality of the \yt~ data, which is known to be
shallower, more inhomogeneous and taken in poorer seeing conditions.
For the case illustrated above, the seeing FWHM in their data is
1.7\arcsec, compared with ours which is around 1.0\arcsec. The poorer
seeing means that a larger photometric aperture must be used, leading
to a noisier measurement of the magnitude.

\subsubsection{Comparison with \yt\ visual morphologies}

We directly compare our visual classifications with the classifications
of \yt, for the 17 EROs in common between our samples.  Though our
classification schemes are different, we can bin both samples coarsely
into disks, bulges and other.  Our classes 2, 5 and 6 are disk
dominated classes and broadly equivalent to their D, D+B and ID
classes. We find 11 objects in common in these three classes.  For
spheroidal designations, we compare our classes 3 and 7 with their B
and BD. Here we find that one object we classify as spheroidal, \yt\ 
class as a disk.  Furthermore, two EROs which they class as bulges, we
classify as disks.  From our merger category (which has no equivalent
in their work), we have two objects, one of which \yt\ class as a
bulge, the other as a disk; and we both agree that a final object is
unidentifiable. Hence, through a very broad comparison with the \yt\ 
classifications, we agree on 12 classifications, disagree on 3 and a
further 2 (our merger category) we are unable to compare.

Next we examine the distributions of morphologies within our ERO
samples.  We restrict our catalogue to $K\leq18.7$, the median
5-$\sigma$ limiting magnitude of \yt, in order to compare the
morphological mixes between the two surveys.  From their visual
classifications, \yt\ found $35\pm5$\% of their EROs to be bulge-like
and $64\pm7$\% to be disk-like with 6\% unclassifiable (where the
errors are simply Poissonian). Our visual classifications divided
coarsely into bulge-like or disk-like classes
(Table~\ref{table:mle_class}) yield $24\pm6$\% bulge-like and
$54\pm10$\% disk-like with 8\% unclassifiable.  Note that here the 14\%
(7/50) of our classifications which do not fit into the simple
bulge-like or disk-like (e.g.\ merger, amorphous) are not included.
Hence, both our surveys find approximately twice as many disks as
bulges at these magnitudes.

\subsubsection{Comparison with MDS automated morphologies}

A more detailed quantitative analysis of galaxy morphology in the MDS
was performed by \citet{1994AJ....108.2362R,1999AJ....118...86R}.  In
outline, they applied a maximum likelihood fitting technique to the
surface brightness profiles of MDS galaxies.  The models they fitted
were axisymmetric de Vaucouleurs and exponential disk profiles.  A
galaxy best fit by the former profile was classed as a bulge and the
latter profile as a disk.  If some combination of the two gave a better
fit then the relative contribution of the two components was output in
the form of a bulge to total luminosity (L$_B$/L$_{Total}$).  The
completeness limit for classification is drawn at $\Xi=1.8$ (where
$\Xi$ is their signal to noise index computed from the signal to noise
of each pixel in an object) which corresponds to $I\sim24.5$ for the
shallowest fields used here. A $\Xi\geq2.0$ (0.5 mags brighter) is
required to class a galaxy as either disk-like or bulge-like (otherwise
sources are just divided into either point or galaxy), and a
$\Xi\gsim$2.4 is required to fit a bulge$+$disk model.

The reliability of this maximum likelihood estimate (MLE) was tested
with the sophisticated and now widely-used two-dimensional surface
brightness fitting algorithm of \citet[][GIM2D]{1998adass...7..108S} by
\citet{2002ApJS..142....1S}. They found that the agreement of
structural properties such as half-light radius and bulge to total
light ratios between the classifiers was very good down to the limit
studied of $I_{814}\leq22$.  A small fraction of objects were found to
have very different bulge fractions as measured by the different
classifiers, and most of these turned out to be peculiar or interacting
systems.  This emphasizes that automated classifiers are unable to deal
robustly with unusual galaxies.

Morphological fractions for our ERO sample based on the MDS MLE
results\footnote{The Medium Deep Survey catalog is based on
  observations with the NASA/ESA Hubble Space Telescope, obtained at
  the Space Telescope Science Institute, which is operated by the
  Association of Universities for Research in Astronomy, Inc., under
  NASA contract NAS5-26555. The Medium-Deep Survey analysis was funded
  by the HST WFPC2 Team and STScI grants GO2684, GO6951, GO7536, and
  GO8384.} are also tabulated in Table~\ref{table:mle_class}. 79 of our
224 EROs are unmatched with any entry in the MDS database. 10 have more
than 1 match within 1\arcsec.  We include these objects and take the
nearest match.  \yt\ found that, for their sample, the MLE morphologies
gave a mix of 50\% disks and 37\% bulges.  Using the same scheme for
our $K\leq18.7$ subsample, we find relative fractions of $32\pm7$\%
disks and $20\pm6$\% bulges with 48\% unclassified.  17 of this
subsample of 50 are not listed in the MDS catalogue
\citep{1994ApJ...437...67G}. Ignoring for the moment the objects
without classifications, the MDS MLE classifications for the \yt\ 
sample find around 40\% more disks than bulges; we find around 60\%
more disks than bulges. The general trends in these two surveys at
bright ($K\leq18.7$) are consistent: visual classifications identify
approximately twice as many disks as bulges and the MDS MLE method
identifies approximately 60\% more disks than bulges for those objects
sufficiently bright to classify.

We have shown via direct comparison that our visual morphologies are in
reasonably good agreement with those of \yt, with a disagreement at the
level of $\lsim20$\%. The relative morphological mixes within both
samples (\yt\ versus our bright subsample) also appear consistent through
both visual morphologies and automated bulge$+$disk decomposition.  Thus
we can now use our full sample to look at evolution across the break in
the ERO number counts into the $K=19$--20 regime.

\begin{table}
\caption{Summary of the coarse morphological classifications for two of
  our ERO samples: the full sample and a subsample cut at $K=18.7$, the median
  limiting magnitude of \yt. Two classification schemes are considered:
  our visual morphologies and automated maximum likelihood estimator
  (MLE) fits. $^\dag$ -- Galaxies in our visual
  classification scheme which do not fit into either the disk or
  spheroid bins, and galaxies which are unclassified by the MDS MLE. See text for details.} 
\label{table:mle_class}
\centering
\begin{tabular}{lcccc}
\hline
Sample & Disks & Spheroids & Other$^\dag$ & Total \\
\hline
\hline
Visual & & & & \\
$K \leq 18.7$ & 27/54\%    &     12/24\%    &     11/22\% &  50  \\
Full  & 84/37\%   &     33/15\%    &     107/48\% &    224  \\
MDS MLE & & & & \\
$K \leq 18.7$ & 16/32\%    &     10/20\%    &     24/48\%     &     50  \\
Full    & 53/24\%    &     33/15\%    &     138/62\% &  224  \\
\hline
\end{tabular}
\end{table}

\subsubsection{Comparison with other morphological studies}

Our morphological fractions contradict the findings of
\citet{2000A&A...364...26M} who found 15\% irregular/disk-like,
$\sim$50--80\% elliptical-like.  Nevertheless, as they themselves note,
their sample is drawn from a heterogeneous collection of archival {\it
  HST} pointings and may not be representative.
\citet{2002MNRAS.337.1282R} also find a high fraction of
bulge-dominated EROs using ground-based $K$-band data ($39\pm11$\%
bulge, $25\pm9$\% exponential disk profile systems) from a subsample of
32 of their EROs with high resolution UFTI data taken in good seeing.
Unfortunately, ground based NIR data is not ideal for morphological
studies, as even in the best seeing conditions, the very nature of the
NIR observing strategy means that multiple, dithered images must be
registered and combined, with the added complexity that the seeing may
be changing between exposures.  Correct propagation of all such
uncertainties through the surface brightness profile fitting procedure
is non-trivial.  Furthermore, \citet{2002MNRAS.337.1282R} report a
possible z$\sim$1 cluster of EROs comprising five bulge systems and two
disks, thus inflating the early-type galaxy fraction.

Our morphological fractions are also in broad agreement with those from
an {\it HST} imaging survey of lensed EROs given by
\citet{2002MNRAS.330....1S} (18\% compact, 50\% irregular/disk-like and
32\% unclassified) using an $(R-K)\geq5.3$ selection criterion.
Although, we caution that $(R-K)\geq5.3$ selection may find different
systems from $(I-K)\geq4.0$ selection (see
\S\ref{sec:ero-sample-diff}).

\section{Results \& Discussion}

\subsection{Evolution of ERO colour distribution}
\label{sec:evolution-ero-colour}
To search for changes in the ERO population across the break in the
number counts, we will take subsamples of the 100 brightest and 100
faintest EROs in the survey, to maintain equivalent uncertainties in
both bins. This corresponds to $K<19.16$ and $K>19.38$ for the bright
and faint samples respectively.  This is a good choice as, although the
position of the break is not precisely defined, it seems to occur
around $K\approx19.0$--19.5.

\begin{figure}
\centering
\includegraphics[width=85mm,clip=t]{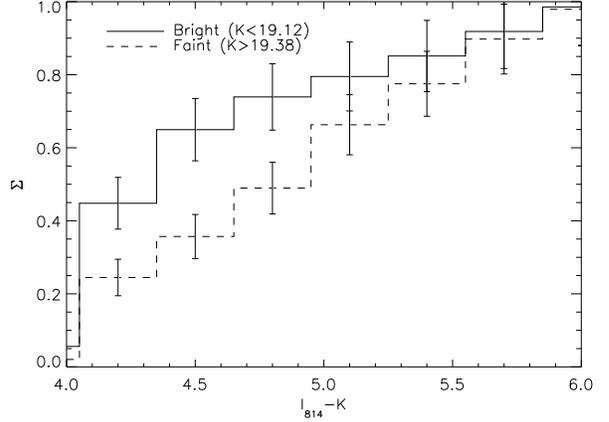}
\caption{A comparison of the cumulative colour distributions of the 100 brightest
  (solid line) and 100 faintest (broken line) EROs in our sample.
  Error bars are based on Poisson statistics in each bin. The fainter
  sample is skewed toward redder colours, indicating that the colour
  distribution evolves, as galaxies pass over the $K\sim19$ break in
  the number counts.  }
\label{fig:diffhists}
\end{figure}

The cumulative colour distributions of the ERO sample divided into
bright and faint $K$-band bins are examined in
Fig.~\ref{fig:diffhists}.  The distributions clearly appear different
such that the fainter sample is skewed toward redder colours (with a
median colour of $(I-K)=4.39\pm0.07$ and $(I-K)=4.95\pm0.09$ for the
bright and faint samples, respectively).  A two-sided KS test reveals
that the probability of the faint colour distribution being drawn from
the same population as the bright sample is $\sim3\times10^{-4}$.  In
this comparison we have also chosen to conservatively treat lower
limits in $(I_{814}-K)$ as detections. Since all but two of the
non-detections occur in the faint bin (see Fig.~\ref{fig:cmd}), objects
in the faint sample can only become redder than the values tested.
This strengthens the conclusion of a reddening of the fainter EROs. If
we completely excise the break region and repeat the test using
$K<$19.0 (81 objects) and $K>$19.5 (86 objects), the trend is
unchanged, and the significance only marginally reduced to
$\sim$6$\times$10$^{-4}$.

Thus, the colours of EROs evolve as they pass over the break in the
cumulative counts, such that a greater fraction of the faint EROs are
redder than their bright counterparts.

\subsection{Evolution of ERO morphologies}

We now investigate whether the changes in the colours of EROs across
the break in their number counts are mirrored by a change in their
morphological mix.

\subsubsection{Morphological number counts of EROs}

\begin{figure}
\centering
\includegraphics[width=85mm,clip=t]{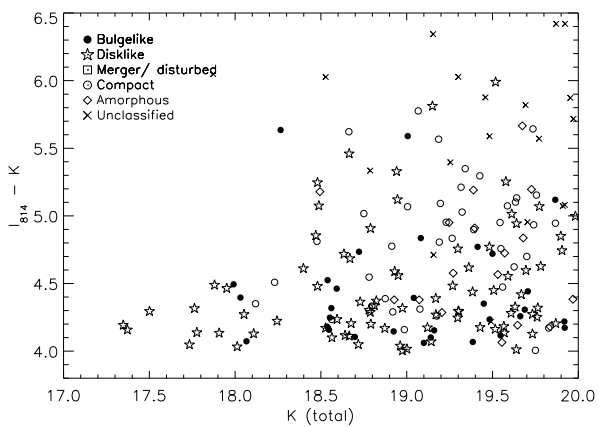}
\caption{The colour-magnitude diagram for the EROs labelled by their visual
  morphologies. Note the dominance of disk-like morphologies in the 
  brightest EROs in our survey, with bulge systems only starting to appear
  at $K>18$. }
\label{fig:morph_cmd}
\end{figure}

\begin{figure*}
\centering
\includegraphics[width=85mm,clip=t]{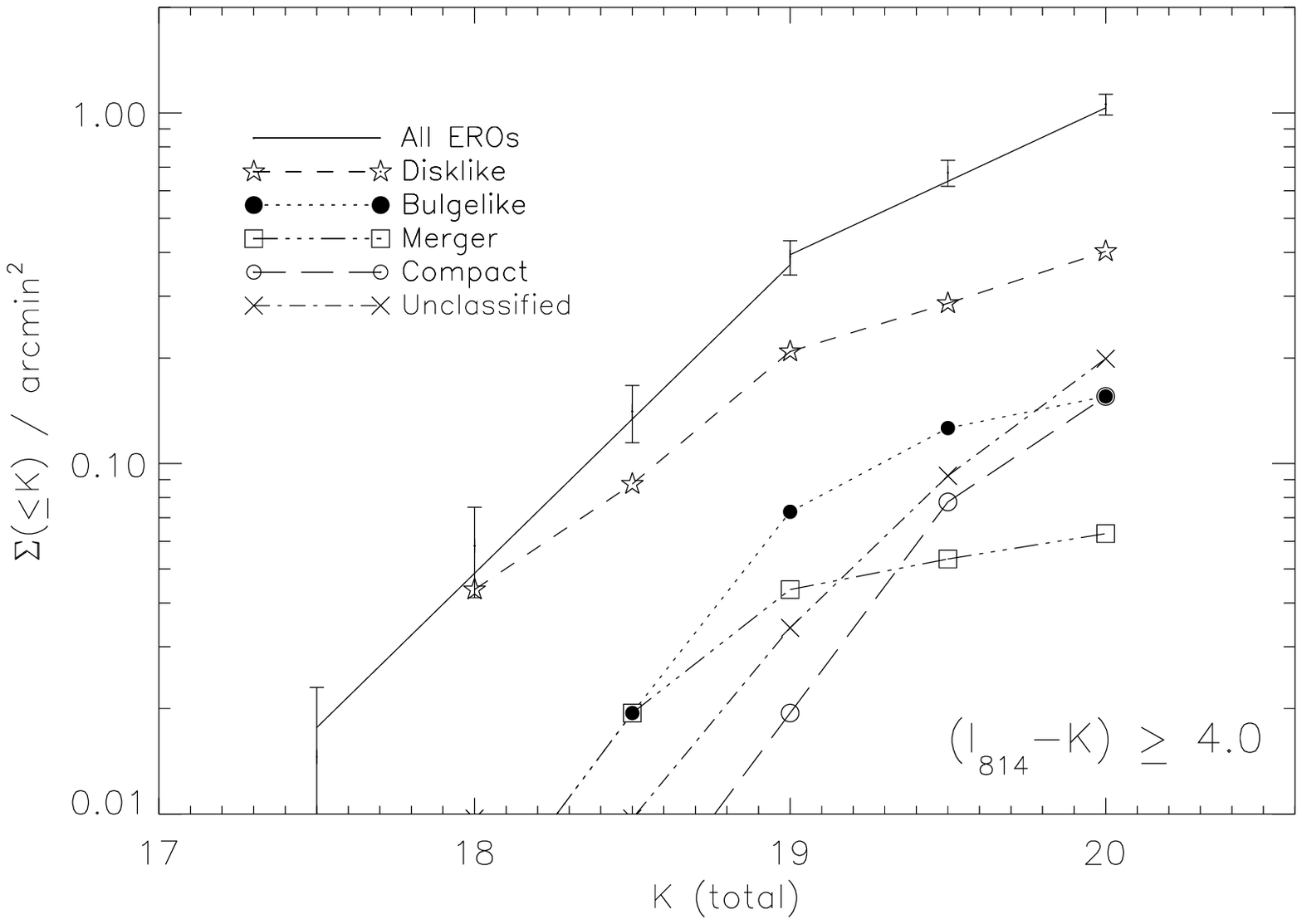}
\includegraphics[width=85mm,clip=t]{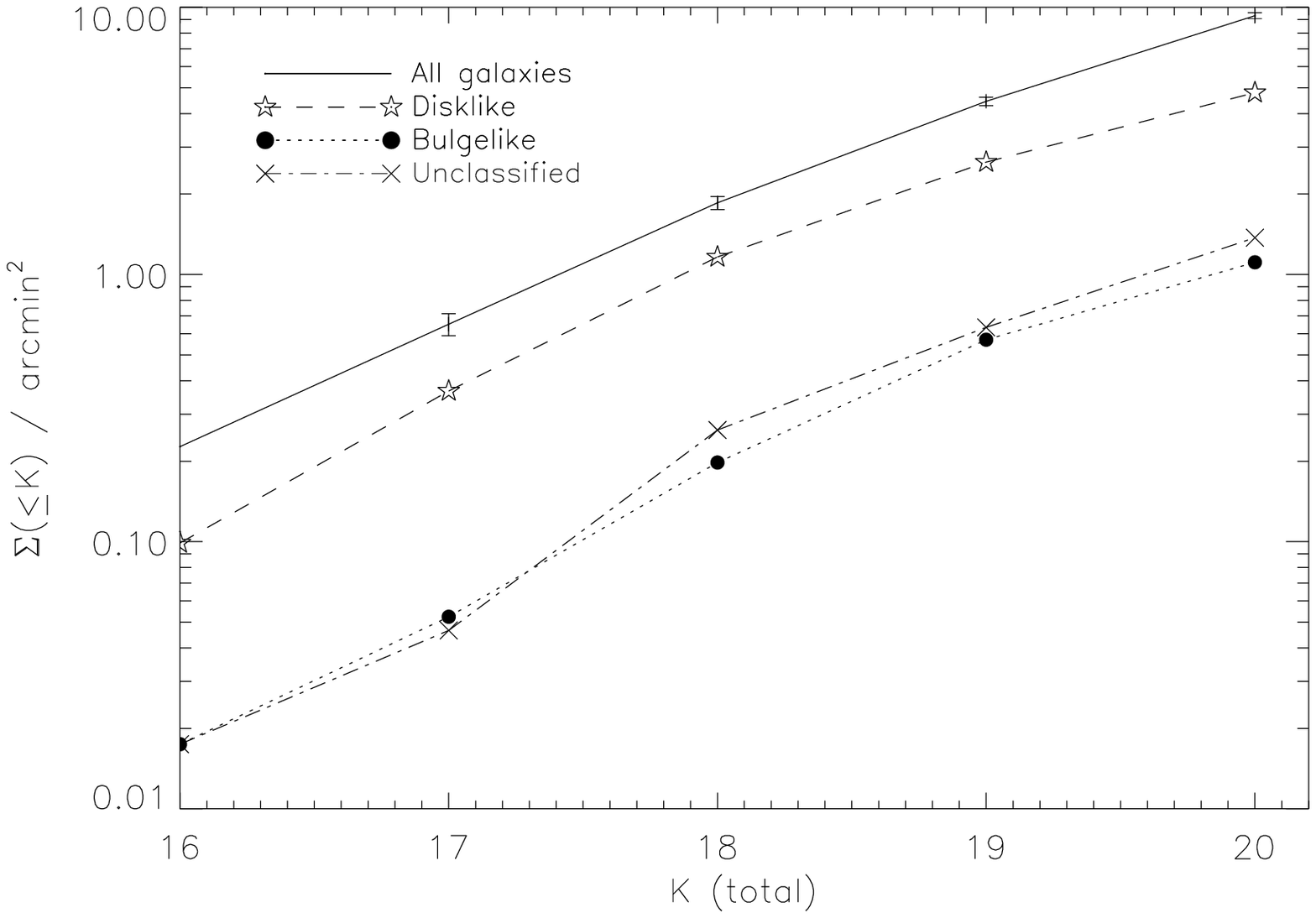}
\caption{ Left panel shows the morphological number counts for our data
  and the double power law fit to the total ERO counts. We do not
  attempt to correct these counts for incompleteness as a function of
  morphological type, as this correction is negligible compared with
  the Poisson uncertainties we assume throughout.  Note that nearly all
  the brightest EROs are disks, and bulges and mergers both appear at
  $K\sim18$. Right panel shows morphological number counts for {\it
    all} K-selected galaxies using MDS MLE classifications from F814W
  data.}
\label{fig:mcounts}
\end{figure*}

We give the distribution of the different morphological subclasses of
EROs on the colour-magnitude plane in
Fig.~\ref{fig:morph_cmd}\footnote{ One particularly bright, red source
  which is unclassified is ERO198 with $(I-K)=6.05\pm0.09$ and
  $K=17.87\pm0.03$.  This source lies on the edge of a halo from a
  bright star. It appears to be a genuine detection, and has an F814W
  counterpart, but its proximity to the star halo {\it may} bias the
  magnitudes measured, as the local sky background possesses a large
  gradient. However, multicolour photometry in additional bands yields
  a plausible photometric redshift of $z \sim$1.7 for this object (see
  \S\ref{sec:phot-redsh}) and it seems to just be a high redshift,
  extreme ERO.  Note that similar objects are seen in the \yt\ data.}.
The main point to note from this plot is that the brightest EROs, which
are responsible for the sharp rise in the number counts at $K\lsim 19$,
are dominated by galaxies with disk-like morphologies. Indeed at
$K\lsim 18$, $\gsim 80$\% of the sample are disk systems.  To better
illustrate the variation in the morphological mix with magnitude we
examine the morphologically-classified number counts in
Fig.~\ref{fig:mcounts}. At $K\sim18$ bulge systems start to appear,
with an even steeper slope than the bright disk population.  Merging
systems also appear at $K\sim18$ and seem to follow the bulge
population.  To quantify these comparisons, we fit power laws to the
morphological subsamples for $K\leq19.0$, where the visual
classifications are reasonably complete ($\sim90\%$), which yields
gradients of $\alpha_{\rm disk}=0.77\pm0.11$; $\alpha_{\rm
  bulge}=1.17\pm0.27$.  For the unclassified (class 0) population:
$\alpha_{\rm unclass}=0.57\pm0.19$ for $K\leq19.0$, and $\alpha_{\rm
  unclass}=0.76\pm0.40$ for $19\leq K\leq20$.  The remaining classes
contain too few objects and/or are not well fit by a power law in this
magnitude range.

Summarising these results in the scheme of Table~\ref{table:visclass}:
to $K=20$, our visual scheme gives $37\pm3$\% disk-like EROs and
$15\pm2$\% bulge-like, with 48\% not fitting easily into either of
these simple classes (19\% of which are unclassified).  Again, we see
around twice as many disks as bulges, but now the unclassified$+$other
fraction is almost twice as high as for the bright subsample. Similarly
if we use the MLE classifications, we find that the ratio of bulge to
disk galaxies is consistent within the uncertainties with that measured
for the bright subsample, but the relative fraction of
unclassifiable$+$other galaxies has doubled.

\subsubsection{Comparison with morphological number counts of all
  K-selected galaxies}

We also fit power laws to the $K$-selected morphological counts for all
galaxies using MDS MLE morphologies from F814W data
(Fig.~\ref{fig:mcounts}).  We find slopes of $\alpha=0.30\pm0.02$ for
the disks and $\alpha=0.35\pm0.02$ for the bulges, in the range
$18.0\leq K\leq20.0$.  The unclassified population follows the bulge
counts closely. The counts for these sub-populations are
substantially shallower than their counterparts in the ERO sample. Our
preferred interpretation of the steepness of the ERO counts is that the
EROs are a high redshift (z$\sim1$, inferred from a passively evolved
M$^\star$ for the bright EROs of K$\sim$18) subsample of the total
$K$-selected galaxy population. We return to the question of redshifts
in \S\ref{sec:phot-redsh}.

\subsubsection{The nature of the bright disk EROs}

To better understand the nature of the bright, ERO disk galaxies
we first measure the average semi-major axes of {\it all} the
well-detected disk systems (as measured by \sex's {\sc a\_image}
parameter).  We find a mean of $<\!A\!> = (0.16\pm0.01)$\arcsec; while
the brighter disk EROs (K$\leq18$) are also systematically bigger:
$<\!A\!> = (0.32\pm0.03)$\arcsec.  Thus these bright (and relatively
blue) disk EROs also appear to be larger than the average
morphologically-classified disks in the ERO population.  Their
elongations range from 1.12 to 1.61 with a median of 1.36, suggesting
that these systems are only moderately inclined ($\sim45^\circ$ to line
of sight).  If these EROs are at z$\sim$1--2, then the median size of
the bright disk sample implies a physical half light
radius\footnote{for all half light radii we use the semi-major axis} of
$\sim3$kpc, compared to $\sim 2$kpc for the full sample of disk
EROs. These sizes are comparable to those seen for similarly luminous
disk galaxies in the local Universe
\citep[e.g.][]{2001A&A...369..421B}. Thus, although these disk EROs are
among the brightest and largest galaxies in the sample, they are
neither too bright nor too large to be conclusively ruled out as lying
at z$>$1.  Indeed, recent work has uncovered large disks (with
half light radii of 5.0--7.5kpc) at redshifts z$=$1.4--3.0
\citep{2003astro.ph..6062L}.

\yt\ claimed that around 40\% of their disk-like EROs (or one third of
all their EROs) showed unusually large, edge on disks and suggest that
these systems are actually lower redshift contaminants to the z$\gsim1$
ERO population.  We find only one very large edge on disk ERO of
comparable size to the one illustrated in \yt's fig.~8 (ERO011 with a
half light radius of 0.45\arcsec, Fig.~\ref{fig:thumbs}). Thus we
suggest that this contamination is slight for the K$\leq20$ ERO
population.

\subsubsection{ERO sub-populations}

Next, we examine the colours and apparent magnitudes of the various
morphological sub-populations of EROs in Table~\ref{table:medcols}. We
consider the disk-like and bulge-like classes, the unclassified
population, the full sample and the amorphous objects \citep[these
should be closest to the irregular EROs studied
by][]{2000A&A...364...26M}.  As would be expected, we see that the
median $K$-magnitude of the unclassified population is fainter (by
$>$0.5 mag) than any other class. The median colour of the faint sample
for the `all' class is seen to be redder by $\sim0.3$ mags than the
faint sample.  This is the colour evolution reported in
\S\ref{sec:evolution-ero-colour}.  The median colours of both the
bright and faint disks and bulges are consistent.
\citet{2000A&A...364...26M} found that irregular EROs were redder than
their bulge EROs.  We see that our amorphous systems are indeed redder
than our bulges, particularly in the bright subsample.  We also note
that in contrast to the full sample, in this case it is the bright
amorphous subsample which is redder than its faint counterpart.
  
Finally, we briefly examine the environment of the EROs, as defined by
the number of $K\leq$20 galaxies within 1 arcmin of each ERO.  This
radius corresponds to a physical size of $\sim$0.5Mpc at z$\sim$1.  We
failed to find any clear trend, with all morphological types inhabiting
similar density environments.

\begin{table*}
\caption{Average properties of ERO sub-populations.  We give the median
colours, and median
K$_{Tot}$ magnitude for various morphological sub-populations. 
Amorphous objects appear significantly redder than
the bulge population (particularly in the bright subsample) as noted by
Moriondo et al.\ (2000). Note also that the faint subsample for all
morphological types is redder than the bright subsample.  This trend is
reversed when just considering the amorphous systems.
Errors are uncertainties on the median from
bootstrap resampling. 
}
\label{table:medcols}
\begin{tabular}{lccccccc}
\hline
Sample & Disks & Bulges & Amorphous & Compact & Merger & Unclassified  & All \\
\hline
\hline
Median$(I-K)$ & & & & & & & \\
$K \leq 19.0$ & $4.30\pm0.05$  &  $4.40\pm0.17$  & $ 5.18\pm 0.31$ &
$ 4.19\pm 0.08$ & $ 4.58\pm 0.22$ &
$6.68\pm0.41$ & $ 4.38\pm 0.05$  \\
$19.0\leq K\leq 20.0$ & $4.41\pm0.10$  &  $4.40\pm0.12$ & $ 4.73\pm0.20$ 
& $ 4.29\pm 0.17$ & $ 4.57\pm 0.13$
& $5.54\pm0.10$ & $ 4.71\pm 0.08$ \\
\hline
\#/\%  & & & & & \\
$K \leq 19.0$ & 43/53.75\% & 15/18.75\% & 2/2.50\% & 
 4/5.00\% & 9/11.25\% &
7/8.75\% & 80/100\% \\
$19.0\leq K\leq 20.0$ & 41/28.47\% & 18/12.50\% & 16/11.11\% 
& 28/19.44\% & 5/3.47\%
& 36/25.00\% & 144/100\% \\
\hline
Median$(K_{Tot})$ & $18.97\pm0.12$ & $19.04\pm0.18$ & $19.56\pm 0.08$ &
$19.54\pm 0.12$ & $18.91\pm 0.18$ &
              $19.63\pm0.10$ & $19.27\pm 0.06$\\
\hline
\end{tabular}
\end{table*}

\subsection{Quantitative evolution of ERO morphologies}

In order to quantify the form of this evolution, we artificially faded
our bright subsample of EROs to the magnitudes of our faint subsample
and statistically compared properties of the artificially faded
population with the true faint population.  The first quantitative
measure we examine is concentration, $C$ (\S\ref{sec:autom-class}).

\subsubsection{Artificially fading the bright ERO population}
\label{sec:artif-fading-bright}
To artificially dim the EROs, each of the 100 $K$-brightest EROs has an
\i814 ~magnitude selected randomly from the observed distribution of
the 100 $K$-faintest EROs.  Each bright ERO was then extracted from the
{\it HST} image, dimmed to its corresponding faint magnitude (keeping a
fixed angular size) and reinserted into a random blank region within
the original F814W image.  Its concentration was then remeasured.  By
repeating this for each of the 100 bright EROs, an artificial faint
sample is created which can be statistically compared with the true
faint sample. By directly fading a bright subsample of our objects, all
the uncertainties associated with measuring properties of the faint
sample are accounted for.  These simulations were repeated 100 times
and the maximum probability of the two samples being drawn from the
same parent population was found to be 0.6\%, with the mean probability
of the 100 simulations being 0.1\%. Thus, we detect evolution in the
concentration indices for EROs across the break in the ERO number
counts.  Note that the angular size change over the expected redshift
range of the objects is unlikely to be responsible: the angular size of
an object only changes by $\sim5\%$ between a redshift of 1 and 2.
Again, this result still holds if we change the bright and faint
samples limits to K$<$19.0 and K$>$19.5, although the maximum
probability is increased to 14\% and the mean is 3\%.

\subsubsection{Evolution in $\mu - C$ space}

The surface brightness -- concentration index ($\mu - C$)
classification plane has been used to quantify galaxy morphology, with
bulge-dominated galaxies typically residing at higher concentration and
greater surface brightnesses than disk-dominated objects (see fig.~1,
\citealt{1994ApJ...432...75A}).

\begin{figure}
\centering
\includegraphics[width=85mm,clip=t]{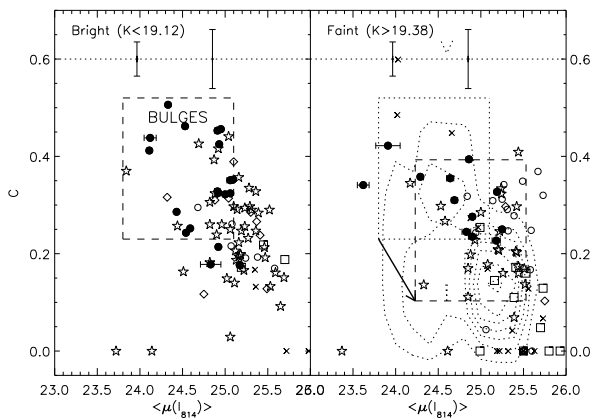}
\caption{The distribution of mean \i814 ~surface brightness ($\mu$) 
  versus concentration index ($C$) for the 100 brightest (left panel)
  and 100 faintest (right panel) EROs in our survey. Symbols denote
  visual morphologies and are the same as for Fig.~\ref{fig:morph_cmd}.
  The uncertainty in $\mu$ is just shown on bulge classes for clarity
  and the typical errors in $C$ estimated from simulations are shown at
  top of each plot. The box in the left panel is the region selected to
  contain a large fraction of visually identified bulges ($\sim 70$\%).
  The arrow in the right panel indicates the translation in $\mu$--$C$
  observed when the bright ERO population is faded to match the faint
  population (as described in text) and the dashed box shows the
  movement of the translated classification box. Contours in the right
  panel show the density of artificially faded galaxies (of all
  morphological types).  It appears that the distribution of the
  simulated faded population extends to lower $ C$ than is seen in the
  real faint sample, suggesting that the low-$C$ galaxies seen in the
  bright sample are absent in the fainter population.}
\label{fig:mu_c}
\end{figure}

Fig.~\ref{fig:mu_c} shows the distributions for our 100 brightest and
100 faintest EROs.  Representative error bars in $C$ determined from
simulations are indicated at the top of each panel and for clarity the
errors in surface brightness are shown only for the bulges.  The spread
of sources in the bright ERO sample is greater than the estimated
errors in both quantities, indicating that these quantitative
observables are measuring a real variation in the morphological
properties within the EROs population, although the distributions for
the visually classified bulges and disks overlap significantly. We can
identify a region in this plane (described below) towards higher
concentration and higher surface brightnesses which contains the bulk
of the visually identified bulge-like population. We take the locations
of our visually classified bulges and disks in $\mu - C$ space as
further support for the reliability of our visual classifications.  In
addition, we note that peculiar (i.e.\ merger and amorphous) systems do
not occupy specific areas of the diagram and are scattered throughout
all the occupied range.

To test the reliability of the apparent evolution on the $\mu - C$
plane, we concentrate on the bulge population and perform a simple
comparison of the expected numbers of faint, bulge-classified EROs. In
the left panel of Fig.~\ref{fig:mu_c} we identify a rectangular region
which contains a large fraction of the visually identified bulges.  We
define the completeness and purity of this box as the fraction of
bulges out of the total number of bulges (in this sample of 100)
located in this box, and the fraction of bulges out of the number of
objects within this box, respectively. For the 35 objects in the box in
the bright subsample we find a completeness of $0.76\pm0.11$ and a
purity of $0.46\pm0.19$.  In the right panel, we move the box by the
mean fading vector determined from the simulations, and recalculate
these values for the faintest 100 EROs, finding a completeness of
$0.42\pm0.16$ and purity of $0.36\pm0.19$ for the 53 objects now in the
box. The box selected is quite large and in the bright sample is
contaminated by at least as many non-bulges as bulges.  However, the
translation of this box and the location of visually classified bulges
in the faint sample shows that the bulge population behaves as
expected, moving along the simulated fading vector, and a comparable
fraction of the faint bulge population is still located within this
region.  We thus conclude that the changing morphological mix between
the bright and faint ERO population is not a result of visual
misclassification. The contours in the right panel denote the density
of artificially faded EROs, the bulk of which lie toward lower values
of $C$ than seen in the real faint sample (as measured in
\S~\ref{sec:artif-fading-bright}).

Hence, it would appear that morphological changes are associated with a
decline in the number of low concentration objects. Galaxies in this
region of the $\mu - C$ plane appear to be disks in the bright sample
\citep[and would be expected to be disks in the local universe,
e.g.][]{1994ApJ...432...75A}.  Thus we interpret the break in the ERO
counts as a real decrease in the number of disk-like galaxies in the
population at $K\gsim 19$.

\subsection{Photometric classification of EROs}
\label{sec:phot-class-eros}
\begin{figure}
\centering
\includegraphics[width=85mm,clip=t]{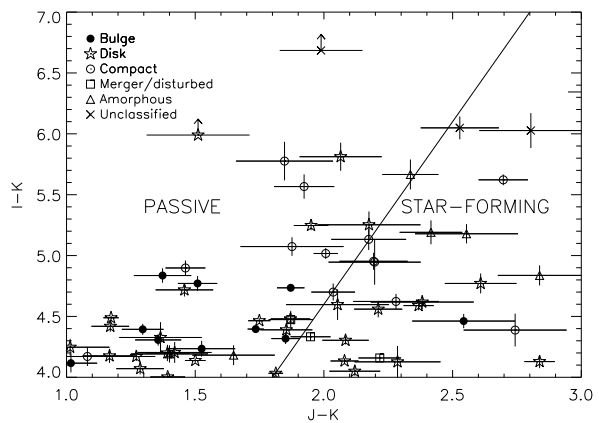}
\caption{The $(I_{814}-K)$ versus $(J-K)$ colour-colour diagram for subsample of our EROs
  with $J$-band photometry.  EROs are labelled with their visual
  classifications. The solid line shows the proposed division between
  photometrically classified passive and star-forming galaxies
  (Pozzetti \& Mannucci 2000). This division is broadly supported by
  our morphological classifications, however a number of EROs with
  disk-like morphologies appear in the lower left corner and hence fall
  in the elliptical category.  As we show in
  \S\ref{sec:phot-class-eros}, a more detailed modelling of the SEDs of
  these galaxies suggests that they are star-forming galaxies which
  this simple photometric classification scheme is failing to correctly
  identify.}
\label{fig:ijk}
\end{figure}

Another possibility for distinguishing evolution in the various
subclasses of EROs is multicolour photometry.
\citet{2000MNRAS.317L..17P} proposed a photometric classification
scheme for separating dusty star-forming SEDs from passive systems in
the $I/J/K$ colour plane.  Using the subsample of our survey with
$J$-band photometry we illustrate this in Fig.~\ref{fig:ijk}.  The solid
line denotes the \citet{2000MNRAS.317L..17P} division between the
bluer, passive and redder, star-forming systems.  We can compare this
classification scheme with the distribution of our visual morphological
classifications to test the extent of the overlap between these two
approaches.  Here we start from the naive assumption that all
morphologically-classified bulge systems will have passive SEDs, while
any galaxy with a disk component is probably an ERO by virtue of a
strong dust component produced by ongoing star formation activity.

51 EROs lie in the passive region of Fig.~\ref{fig:ijk}, and 21 in the
star-forming region. Firstly we quantify the fractions of
photometrically classified objects in each morphological class.  Of the
30 disks, 21 (70\%) are classified as having star-forming SEDs, with
only 9 (30\%) falling into the passive region.  Equally, 10 (91\%) of
the 11 bulges are classed as passive and 11 (78.6\%) of the 14 compact
sources have colours consistent with passive SEDs.  For the remaining
classes, 2 (66\%) of the 3 mergers; 3 (50\%) of the 6 amorphous and 6
(60\%) of the 10 unclassified objects have SEDs classed as
star-forming.

Thus the photometric classifications crudely support our prejudices
based on the morphologies of the EROs: most disk systems have
star-forming SEDs and most bulge systems have passive colours; indeed
we find only one morphologically-classified bulge lies in the
star-forming region. The main discrepancy comes from a number of disks
falling into the passive region of the plot, although these are mostly
relatively blue in $(I_{814}-K)$, with $(I_{814}-K)\lsim 4.5$.  Closer
inspection of the morphologies of these EROs does not show strong
evidence of them being early-type disk galaxies (which could be
expected to show passive SEDs).

Another clear trend from this comparison is that a similar fraction of
compact sources are classed as passive to that seen for the bulge
population. This is consistent with the idea that the compact class
comprises bulge systems which are too faint to be unambiguously
classified as such.  The amorphous, merging and unclassified samples
show much higher proportions of sources with star-forming SEDs,
suggesting that these indeed represent morphologically more complex
systems, rather than misclassified faint bulges.

Hence using our visual morphologies we find good agreement with the
photometric scheme of \citet{2000MNRAS.317L..17P} for the higher signal
to noise bulge-classified galaxies and less agreement for the
disk-classified ERO sample. We use this to infer the probable
morphological mix of the lower signal to noise amorphous and compact
EROs, associating the amorphous systems with star-forming and the
compact systems as mostly passive.

S02 used the equivalent $R/J/K$ colour-colour diagram to analyse the
colours of radio-detected and undetected EROs.  They found a similar
level of agreement, in that the EROs expected on the basis of their
radio emission to be star-forming mainly fell into the star-forming
region of the diagram, with some cross-contamination.
\citet{2002MNRAS.330....1S} also used this colour-colour diagram to
examine the location of their {\it HST} imaged EROs.  They found that
all their objects with compact morphologies lay within the passive
region. The size of their $J$-band photometric errors prevented them
from drawing conclusions from their irregular objects.  An alternative
but similar colour selection criterion in $(I-J)$--$(J-K)$
colour-colour space was used by \citet{2002ApJ...577L..83W} to study
the colours of EROs with sub-mm observations. They found that objects
in the star-forming region of the diagram were indeed responsible for
the bulk of the 850$\micron$ emission detected from their coadded
observations, indicative of dusty star-formation activity.
\citet{2003astro.ph..3206R} adopted a similar scheme using galaxy
tracks plotted in $I/J/H/K$ colour space to attempt to separate the
classes.  They noted, as with these other works, that a large fraction
of the EROs lie close to the dividing line between classes.  This could
be symptomatic of the size of the photometric errors, or indicate that
most of this class of object do indeed possess SEDs intermediate
between dusty star-forming and passive stellar populations.

\subsection{Photometric Redshifts}
\label{sec:phot-redsh}
22 of our ERO sample have good detections in four photometric bands:
$R_{606}$ and $I_{814}$ from the MDS observations and $J$ and $K$ from
our INGRID survey.  The MDS F606W observations of these fields were
retrieved from the ST-ECF archive and reduced in a similar manner to
the F814W images (\S\ref{sec:it-hst-imaging}).  For these galaxies it
is possible to investigate the properties of their SEDs in more detail.
However, the apparent colours are affected both by their intrinsic SED
and their redshift and we are therefore required to fit for both of
these variables.  Hence, we have used a photometric redshift code, {\sc
  hyperz} \citep{2000A&A...363..476B}, to study these objects.  We
adopted a procedure similar to S02.  We attempt to fit, in turn, a
dusty star-forming and an evolved SED to each object.  We use a single
star-formation history with an e-folding time-scale of $\tau =$1Gyr, a
Miller-Scalo IMF and solar metallicity.  The dusty SED is allowed
reddening values in the range $A_V = 1 - 6$ \citep[consistent with
values estimated by][]{2002A&A...381L..68C}, and the evolved SED must
have a reddening value $A_V\leq0.2$. Redshifts in the range 0 to 4 are
considered. We list the photometric data and the derived properties for
this sample in Table~\ref{table:hypz}.  In this analysis we retain the
F606W passband, rather than introducing any uncertainty by transforming
it to another, more common, filter system.  We consider only the EROs
with {\it detections} in all four passbands, rather than including
limits as well, as using three or fewer detections leads to very poorly
constrained SEDs. If this selection introduces any bias, it will be
that requiring detections in the optical passbands preferentially
selects bluer galaxies. However, we are primarily interested in the
relation between SED classifications and colour/morphological
classifications, rather than constructing a fair sample.
 
From the 22 EROs, 14 could be fit using {\sc hyperz}; the remaining 8
are typically the reddest objects in (F606W$-K)$. All but one have
(F606W$-K)\gsim7$.  The median redshift of our sample with photometric
redshifts, which has a median magnitude of $K=19.43\pm0.19$, is
z$=1.20\pm0.22$ with all but two EROs predicted to lie in the range 0.7
to 1.8. \citet{2002A&A...381L..68C} found a median redshift of
$<$z$>=1.1\pm0.2$ with galaxies in the range $z=0.7$--1.4, for
$(R-K)\geq5.0$, $K\leq19.2$ EROs, using optical spectroscopy.  Thus our
sample spans the same redshift range as that of
\citet{2002A&A...381L..68C}, but with a higher upper limit, as might be
expected from the fainter magnitude range probed here.

For the fitted EROs we find 11/14 have disk-like morphologies, with a
further 2/14 either amorphous or LSB and one unclassified (the high
fraction of morphologically-classified EROs in this relatively faint
$K$-band subsample simply reflects our selection of EROs with strong
optical detections).

The photometric classifications give a dusty star-forming SED as the
best fit for all but two cases which favour evolved SEDs.
Interestingly, half of these EROs (7/14) with dusty star-forming SEDs
lie in the passive region of the $(I_{814}-K)$--$(J-K)$ plane
(Fig.~\ref{fig:ijk}), with many of the EROs occupying the
$(I_{814}-K)\sim4.0$ region in the lower left corner of the diagram.
This may reflect our selection, where by requiring the EROs be red in
$(I-K)$ but sufficiently blue to be well detected in $I_{814}$ and
F606W, we select unusual systems, possessing strong UV upturns and
possibly a mixture of young and evolved stellar populations.  We note
that of the blue disks in the passive region, 29\% (2/7, or 2/9 if we
include two borderline passive cases) are best fitted with evolved
SEDs, suggesting that these are indeed passive disk systems. 

One of these EROs (\#226, an amorphous object) is best-fitted as a
low-luminosity, dusty system at low redshift (z$\lsim$0.4), but this is
only marginally favoured over an evolved SED at z$=$1.2. Overall, it
seems unlikely that a large fraction of our EROs are at such low
redshifts (z$\sim$0.4). Moreover, such systems are not seen in the
brighter \citet{2002A&A...381L..68C} spectroscopic sample.

\subsection{High-redshift EROs}

One of the EROs in the photometric analysis in the previous section has
a much higher estimated redshift than the remainder of the sample.
This ERO is \#158 at $z_{phot}=3.39^{+0.11}_{-0.26}$. This source is
one of only two EROs in this subsample with $(J-K)\geq 2.3$.  Van
Dokkum et al.\ 2003 (see also \citealt{2003ApJ...587L..79F}) have
recently shown through optical spectroscopy that selecting galaxies
with $(J-K)\geq2.3$ efficiently selects objects with a prominent
optical break (either the 3625\AA~Balmer break or the 4000\AA~Ca{\sc
  ii} H$+$K break) at $z\geq 2$. We can use all of our MDS fields with
$I/J/K$ photometry (81 arcmin$^2$) to examine the nature of this class
of ERO. Adopting this selection, we find 24 EROs with $(J-K)\geq 2.3$,
or a surface density of $(0.30\pm0.06)$ arcmin$^{-2}$ (Poisson error)
to $K\leq$20.  Van Dokkum et al.\ 2003 only give the surface density at
$K=21$ ($1.09^{+0.20}_{-0.16}$) and so we are unable to compare with
their numbers, except to note that our surface density is lower to this
brighter magnitude limit, and that there are very large field to field
variations in this class of ERO \citep[there are {\it no}
$(J-K)\geq2.3$ EROs in the HDFS,][]{2003ApJ...587L..83V}.  The
morphological mix of these galaxies is $20\pm9\%$ disk-like, $4\pm4\%$
bulge-like, $17\pm7\%$ amorphous, $8\pm6\%$ merger, $13\pm7\%$ compact
and $38\pm13\%$ unclassifiable.  The fraction of bulges relative to
disks in the sample is lower than for the full $(I-K)\geq4.0$ ERO
sample, although just compatible within the 1-$\sigma$ Poisson errors.
The fraction of peculiar (not disk-like or bulge-like) EROs is higher
than for the full sample. The magnitudes of these $(J-K)\geq 2.3$
selected EROs span the full range of the $(I-K)$ ERO sample.  It is
interesting to note the lack of bulge-like systems in this perhaps
higher redshift sample, and also the absence of passive spectra from
\citet{2003ApJ...587L..83V}'s sample.  This could be an indication that
at these redshifts (z$\gsim2$) the formation epoch of early type
galaxies is being approached \citep[as suggested by complimentary
studies of cluster ellipticals, e.g.][]{1998ApJ...492..461S}.

\begin{table*}
\centering
\caption{Photometric properties of EROs with F606W, $I_{814}$, $J$ and $K$
  data. In the SED fitting procedure, the raw F606W magnitudes have
  been used directly instead of attempting to convert to
  $R_{606}$. $^a$ -- best fitting SED type: D -- dusty; E -- evolved. $^b$ -- PM00 shows the classification based on the Pozzetti
  \& Mannucci scheme. P -- passive, S -- star-forming. Square brackets
  indicate systems close to the dividing line, whose errors allow them
  into the other class. $^c$ -- the 99\% confidence
  interval for the photometric redshift. $^d$ -- morphology
  refers to visual classification.  }
\label{table:hypz}
\begin{tabular}{rcccccccccp{3cm}}
\hline
ID   &  (F606W$-K$) & $(I_{814}-K)$ & $(J-K)$ & $K$ & SED$^a$ & PM00$^b$ & $z_{phot}$ &
$z_{phot}$ & Morph$^d$ & Comments\\
     &          &  &             &     &    &            & $(99\%)^c$ &
     & & (visual morphology)\\

\hline
001 &  6.93$\pm$ 0.14 &  4.60$\pm$ 0.13 &  2.05$\pm$ 0.22 & 19.69  &
D & [S]  & 1.20       &  0.28--1.49  &2 & \\
002 &  6.62$\pm$ 0.13 &  4.33$\pm$ 0.13 &  1.37$\pm$ 0.16 & 19.63  &
D & P  & 1.18       &  0.93--1.51  &2 & early-type?\\
158 &  5.78$\pm$ 0.06 &  4.13$\pm$ 0.05 &  2.84$\pm$ 0.06 & 18.11  &
D & S & 3.86       &  3.20--4.00  &6 & edge-on late-type disk \\
167 &  6.04$\pm$ 0.14 &  4.13$\pm$ 0.13 &  2.29$\pm$ 0.17 & 19.73  &
D & P & 0.71       &  0.15--1.26  &6 & small edge-on disk 1\arcsec/ late-type \\
197 &  5.66$\pm$ 0.09 &  4.20$\pm$ 0.09 &  1.42$\pm$ 0.14 & 19.87  &
D & S & 1.49       &  1.26--1.79  &2 & \\
198 &  8.78$\pm$ 0.23 &  6.05$\pm$ 0.09 &  2.53$\pm$ 0.15 & 17.87  &
D & [S] & 1.71       &  0.83--1.93  &0 & \\
204 &  7.27$\pm$ 0.17 &  5.25$\pm$ 0.11 &  2.17$\pm$ 0.23 & 19.58  &
D & [P] & 1.00       &  0.95--1.9  &2 & \\
206 &  7.27$\pm$ 0.05 &  4.30$\pm$ 0.03 &  2.08$\pm$ 0.09 & 18.78  &
D & S & 0.89       &  0.58--1.18  &6 & edge-on Scd/Sdm 2.5\arcsec \\
209 &  8.42$\pm$ 0.10 &  5.25$\pm$ 0.04 &  1.95$\pm$ 0.07 & 18.48  &
D & [P] & 1.49       &  1.44--1.66  &2 & (near chip edge) \\
224 &  5.77$\pm$ 0.07 &  4.18$\pm$ 0.06 &  1.17$\pm$ 0.10 & 19.43  &
D & P & 1.49   &  1.34--1.61  &6 & edge-on disk with tidal tail/cmp 2" \\
225 &  6.18$\pm$ 0.05 &  4.17$\pm$ 0.04 &  1.27$\pm$ 0.07 & 18.53  &
E & P & 1.09       &  1.00--1.10  &6 & face-on asymm disk 1.5" \\
226 &  5.84$\pm$ 0.08 &  4.19$\pm$ 0.07 &  1.40$\pm$ 0.15 & 19.65  &
D & P & 0.10       &  0.00--0.42  &9 & \\
227 &  5.97$\pm$ 0.07 &  4.72$\pm$ 0.05 &  1.46$\pm$ 0.11 & 18.63  &
D & P & 1.79       &  1.51--2.00  &5 & face-on disk or clumpy LSB\\
229 &  6.09$\pm$ 0.06 &  4.07$\pm$ 0.05 &  1.29$\pm$ 0.09 & 19.14  &
E & P & 1.06       &  0.97--1.09  &6 & late-type slightly inclined disk 2\arcsec \\
\hline
\end{tabular}
\end{table*}

\subsection{ERO sample differences in different photometric passbands}
\label{sec:ero-sample-diff}
A range of different photometric criteria have been used to identify
`Extremely Red Objects'. How do the different criteria affect the mix
of objects selected?  The ERO number counts for samples selected by a
variety of definitions (e.g.\ $(R-K)\geq5.3$, $(I-H)\geq3.0$) are in
broad agreement (e.g.\ fig.~7 of \yt), and in addition the
$K\approx19$--20 break in the counts is seen in sufficiently deep data
selected by $(I_{814}-K)$, $(I-H)$, and $(R_{702}-K)$ \citep[this
work;][respectively]{2001ApJ...560L.131M,2002MNRAS.330....1S}. However,
a direct comparison of the details of the samples such as the relative
morphological mixes of EROs using different colour selection has yet to
be addressed.

\begin{figure}
\centering \includegraphics[width=85mm,clip=t]{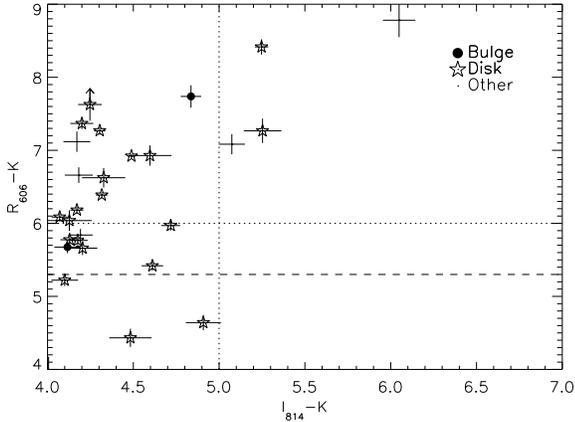}
\caption{The $(R_{606}-K)$--$(I_{814}-K)$ colour-colour diagram for our ERO sample (selected
  by $(I_{814}-K)\geq 4.0$) with available F606W photometry. The dashed
  line shows the typical $(R_{606}-K)\geq 5.3$ selection bound for
  $R$-selected EROs, and the redder $(R_{606}-K)\geq 6.0$ limit
  discussed in the text.  Points are labelled according to their visual
  morphologies in broad classes -- disk-like, bulge-like, or
  unclassified. Vertical error bars represent only the error in the
  F606W photometry and do not account for the uncertainty in the
  transformation to $R_{606}$. This suggests that the $(I_{814}-K)\geq
  4.0$ selection criterion is less stringent than $(R_{606}-K)\geq
  5.3$. }
\label{fig:rik}
\end{figure}

\begin{figure*}
\centering \includegraphics[width=180mm,clip=t]{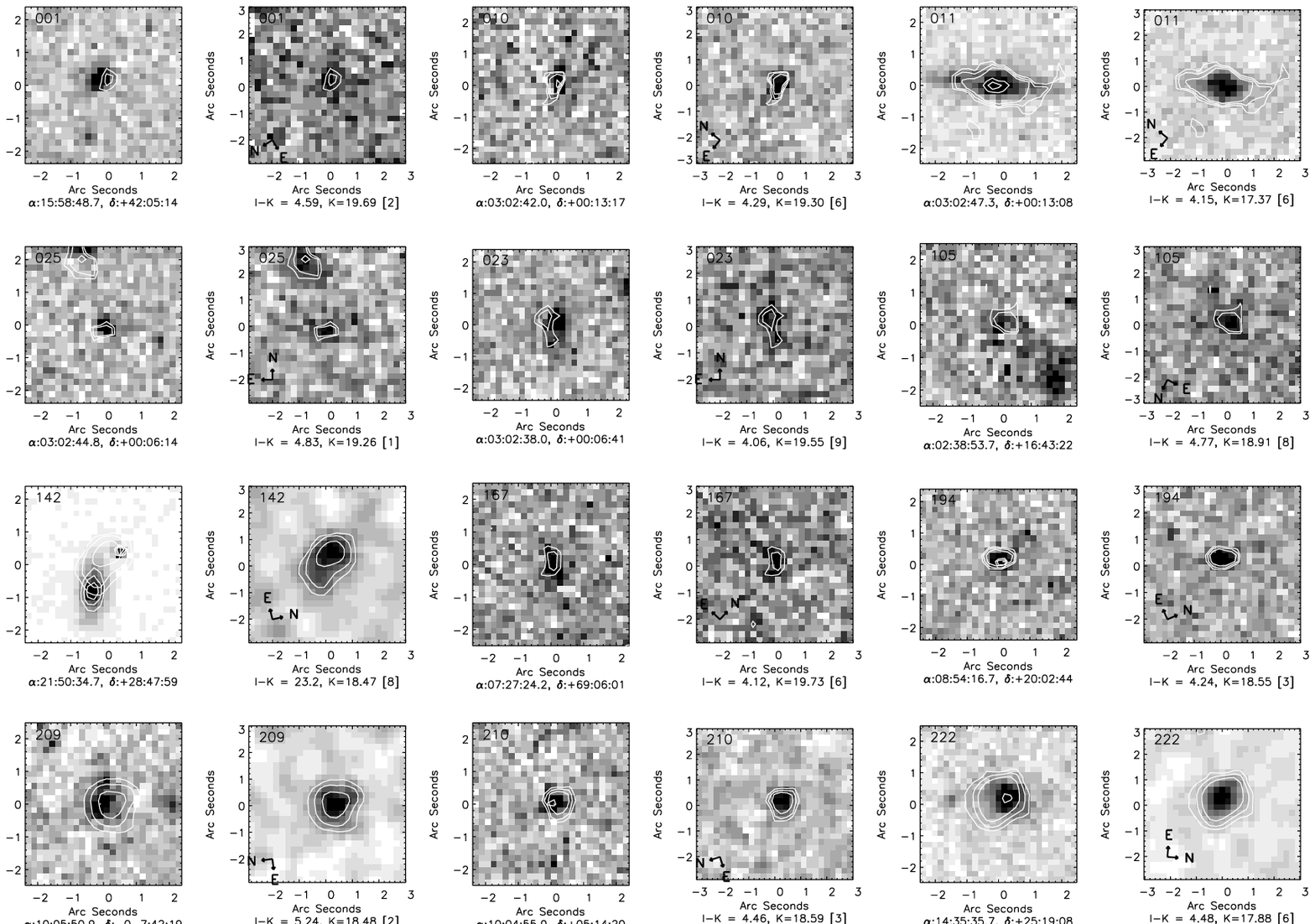}
\caption{Thumbnail images of a examples of EROs.  Each set of images
  comprises an F814W and \ks-band pair; ERO ID number is given in upper
  left.  Left panel (labelled with R.A.  and Dec.) is the non-resampled
  F814W image, rebinned 2$\times$2 in its original orientation; right
  panel is \ks-band image, rotated to match the F814W orientation, with
  the vector indicating the orientation on sky.  The label gives the
  $(I-K)$ colour, $K$ magnitude and visual morphological class in
  square brackets. The contours show the \ks-band isophotes. ERO11 is
  the largest disk in our sample. The \ks-band images appear overly
  smoothed as they have been resampled to match the (non-resampled)
  \i814 ~image in this illustration, but are not resampled in our
  analysis.  }
\label{fig:thumbs}
\end{figure*}

The difference between the $(R-K)$ and $(I-K)$ colour cuts can be
examined for a subsample of our ERO sample with F606W data.  F606W can
be approximately transformed to the Cousins $R$ passband using $R_{606}
\approx\,$F606W$_{Vega} - 0.37($F606W$-$F814W$)_{Vega}$
\citep{metcalfe00}. This results in 27 EROs with measurable $R_{606}$
photometry (Fig.~\ref{fig:rik}).  Only three of the $(I_{814}-K)$
selected EROs are bluer in $(R_{606}-K)$ than the typical selection
criterion of $(R_{606}-K)\geq5.3$. One of these sources is very close
to the selection boundary and would be found using the
$(R_{606}-K)\geq5.0$ criterion \citep[e.g.][]{2002A&A...381L..68C}, and
may even be found with the former criterion, given the uncertainty in
the F606W to $R_{606}$ transformation. The remaining two EROs have
$(R_{606}-K)\sim4.0$ and would not.  Thus, $\approx$90\% of our EROs
would also be selected in an $(R_{606}-K)$ ERO survey (in agreement
with S02 who found a value of 93\%), suggesting that
$(I_{814}-K)\geq4.0$ is a less stringent ERO requirement than
$(R_{606}-K)\geq5.3$.  Furthermore, for this admittedly small sample,
all the galaxies dropping out of the ERO category when selected in
$(R-K)$ are disks, which supports the suggestion by \yt\ that $(I-K)$
selected ERO surveys may preferentially include disks. To
quantitatively compare the different morphological mixes found by the
two techniques requires $R$ and $I$ band data and {\it HST}
morphologies on the same regions of sky (e.g.\ Gilbank et al.\ in
prep).

We also performed visual classifications of the EROs in F606W to
compare with the morphologies derived from the F814W images. We find
for the 27 EROs with sufficient signal for F606W photometry: 8 of these
are too faint to morphologically classify; 10 of these are disk-like in
F606W and were also disk-like in F814W; 2 galaxies were compact and
disk-like in F814W and were simply compact in F606W; 5 galaxies appear
LSB and disturbed in F606W whereas these systems were just classed as
disk-like in F814W; one galaxy appears amorphous in F606W and disk-like
in F814W. One ERO classed as compact and disk-like in F606W was classed
as compact and symmetrical in F814W (nominally a bulge class). This
represents the only discrepancy between our notional bulge and disk
classes, but the difference is slight as in both passbands the object
is compact and symmetric. Thus, the agreement between the F814W and
F606W visual morphologies is very good.  There is a tendency for the
disks in the latter to exhibit more disturbed morphologies, as would be
expected as the F606W probes further into the extreme ultraviolet, in
each EROs' rest-frame. This agreement between passbands is reassuring,
as F606W at z$\sim 1.0$ corresponds to the same rest-frame wavelength
as F814W at z$\sim 1.5$ -- a plausible redshift spread for the ERO
sample.  For this small subsample, we only find one (marginal) bulge --
disk misclassification, whereas we assume Poisson uncertainties on all
our measured fractions throughout this work.

\subsection{Composition of the ERO population}

Our visual and quantitative analysis of the morphological mix of our
ERO catalogue suggests that the ERO population at $K\leq 19$ is
dominated by galaxies with disk-like morphologies (54\% of the total
sample), with EROs with bulge-like or compact morphologies making up
around a further quarter and a modest fraction ($\sim 15$\%) of EROs
with disturbed, merging or amorphous morphologies.  As we have shown,
this distribution is similar to that found by previous studies (\yt;
\citealt{2002MNRAS.330....1S}).  In contrast, our faint ERO subsample,
with $K=19$--20, has a lower fraction of disk-like EROs, down to less
than a third, with the fractions of the other classes effectively
unchanged and the decline in the disk-like EROs being made up for by a
corresponding increase in the fraction of those with amorphous, merger
or unclassified morphologies.  Our quantitative tests of this evolution
confirm that the apparent decline in the disk-like fraction is real and
is not due to our inability to classify faint examples of this
population.

We can now ask what limits we can place on the proportion of the ERO
population arising from galaxies which are red due to dust-obscuration
and what fraction comes from evolved, passive systems.  We use our
visual classifications to divide our sample into star-forming or
passive EROs by taking extreme values in the following way.  We assume
that the minimum fraction of star-forming EROs is given by all the
disk-like galaxies \citep[although note that passive disks are also
likely, \S~\ref{sec:phot-redsh};][]{2002MNRAS.333L..16S}, and the
maximum star-forming fraction is given by all the galaxies which could
plausibly harbour star-formation (including the disks, amorphous,
mergers and {\it all} the unclassified objects). In the same way we
associate all bulge-like EROs with passive systems and add all compact
and unclassified sources to this to generate an upper limit.  We thus
find 54--77\% star-forming and 19--31\% passive EROs in the bright
subsample ($K\leq19.0$), and 29--71\% star-forming and 13--58\% passive
in the faint sample ($19.0\leq K\leq20.0$). For the full sample these
numbers are 38--73\% and 15--48\% for star-forming and passive EROs,
respectively.  S02 (based on an ERO sample selected using
$(R-K)\geq5.3$) used deep radio data to estimate the fraction of star
forming EROs, finding a value of $\sim45$\%, to $K=20.5$.  This is
compatible with our estimate based on morphologies. Furthermore, they
used photometric SED fitting to assess the fraction of dusty galaxies,
placed a firm lower limit of $\geq30$\%, and estimated that possibly
$60\pm15$\% of this ERO population could be dusty star-forming
galaxies.  Again, this
does not contradict our findings. 

In the discussion above, it is the increasing fraction of
morphologically unclassified EROs in the faintest samples which
produces the greatest uncertainty in the evolutionary trends we derive.
These galaxies make up a quarter of the $K=19$--20 sample and appear to
also be responsible for the reddening of the whole ERO population at
$K\gsim 19$, especially when balanced against the declining fraction of
bluer, disk-like EROs (Table~\ref{table:medcols}).  It is tempting to
associate this population with the counterparts to the dusty, active
sources selected in the sub-mm waveband
\citep[e.g.][]{2002MNRAS.331..495S, 2002ApJ...577L..83W}. However, the
confirmation of this will have to await complimentary multiwavelength
and morphological studies, with the latter either benefiting from the
higher red sensitivity of the new ACS camera on-board {\it HST} and the
revival of the NICMOS near-infrared imaging capability of {\it HST}.

\subsection{Cosmological significance of EROs}

Assuming the median photometric redshift ($z\sim1.5$) for our sample,
the break around $K=19$ corresponds to a luminosity of $M_V\sim-19.7$.
For the star-forming population, we correct this value for reddening to
give $M_V\sim-22.1$ \citep[assuming
$A_V\sim2.4$,][]{2002A&A...381L..68C}.  This unobscured luminosity
corresponds approximately to an $L_V^\star$ galaxy today.
Conservatively assuming no further star-formation activity in these
galaxies between $z\sim 1.5$ and the present day (and no substantial
merging), these galaxies would actually correspond to sub-$L_V^\star$
galaxies at $z=0$.  Similarly, if we adopt passive evolution models for
the passive/bulge population, then $K=19$ at $z\sim 1.5$ corresponds to
an $L_V^\star$ elliptical today.

Taking the range of our photometric redshifts ($0.7\leq z \leq1.8$,
ignoring the two outliers) we derive a comoving volume for our survey
of $\sim2\times10^5\,$Mpc$^{-3}$. This gives a space density for our
full ERO sample of $\sim1\times10^{-3}\,$Mpc$^{-3}$. This is only a
crude estimate since we do not know the true redshift distributions of
the galaxies, and furthermore, the redshift distributions of the dusty
star-forming and passive EROs are unlikely to be the same. Thus we use
only an order of magnitude estimate. This number density agrees to this
level with that derived locally from SDSS$+$2MASS data for $\geq
L^\star$ galaxies \citep{2003astro.ph..2543B}.  Thus, the ERO
population could plausibly account for a sizable fraction ($\gsim
10\%$, and potentially all) of the stars seen in luminous ($\sim
L^\star$) galaxies locally.

A more detailed comparison of the ERO counts with a variety of
evolution models was undertaken by \citet{2002MNRAS.330....1S}, who
compared their ERO counts with Pure Luminosity Evolution (PLE) and
semi-analytic hierarchical galaxy formation models.  The only model to
correctly reproduce the cumulative ERO number counts was one in which
the ERO population was modelled as a single population of passive
galaxies.  This model now clearly overestimates the number of passive
EROs as we have been shown that a sizable fraction of the EROs are not
simple passive systems.  The more realistic models examined by
\citet{2002MNRAS.330....1S} -- PLE with a mixture of galaxy types and
semi-analytic galaxy formation models -- both under-predicted the number of
EROs by around an order of magnitude.  A comparison with these models
is beyond the scope of this paper and will be presented in future
work.  However, we note briefly that more recent semi-analytic models,
which include bursts of star formation at high redshift absent in the
\citet{2000MNRAS.319..168C} reference model, now reproduce the number
counts for $(I-K)\geq4.0$ EROs (C.\ Baugh, priv.\ comm.).
Unfortunately, the numbers of objects in redder subsamples (e.g.\ 
$(I-K)\geq5.0$) are still underestimated by around an order of
magnitude. The requirement of these starbursts at higher redshift
(which is also indicated by the presence of luminous, passive galaxies
at z$\,\sim1$) is consistent with the interpretation of SCUBA sources as
protoelliptical galaxies in the process of formation \citep[e.g.
][]{2002MNRAS.331..495S}.

\section{Conclusions}

We have compiled a new catalogue of $K$-selected Extremely Red Objects
in regions with deep {\it HST} $I_{814}$-band observations. We apply a
number of tests to examine the nature of the ERO population and
investigate the origin of the turnover in the ERO number counts around
$K\sim19$.  The main aim of our work is to study the morphologies of
our EROs using the exquisite resolution of the {\it HST}. However, for
a subsample of our dataset with multicolour data we can also utilise
SED-fitting as a classification tool.  Hence, we were able to examine
the commonly applied hypothesis that the passive population defined by
multicolour photometry (or equally spectral line diagnostics) is
associated with bulge-dominated morphologies; and that EROs with the
colours of dusty star-forming galaxies have disk-like/irregular
morphologies.

1. We find a surface density of $(I_{814}-K)\geq4.0$ EROs
$(1.14\pm0.08)$ arcmin$^{-2}$ at $K=20.0$.  The number count slope
flattens from $\alpha=0.88\pm0.09$ for $K\leq19.0$ to
$\alpha=0.42\pm0.19$ for $19.0\leq K\leq20.0$, in good agreement with
other surveys in the literature.

2. This turnover in the number counts is associated with a reddening of
the faint ERO population, from a median colour of
$(I_{814}-K)=4.38\pm0.05$ at $K\leq19.0$ to $(I_{814}-K)=4.71\pm0.08$
at 19.0$\leq K \leq$20.0.

3. Our visual morphological fractions at the bright end (K$\leq$19.0) are in
good agreement with \citet{2003ApJ...586..765Y}, and we extend this
sample to fainter depths, finding a mix of 35\% disk-like, 15\%
disturbed/irregular, 30\% spheroidal or compact and 20\%
unclassifiable.

4. Using quantitative measures, we find that the concentrations of EROs
evolves across the break in their number counts, such that the fraction
of disk-like galaxies declines.  

5. We find on the basis of $I/J/K$ colours, using the scheme of
\citet{2000MNRAS.317L..17P}, that most disk-like and amorphous galaxies
are associated with dusty star-formation activity, and most bulge-like
and compact sources have the colours of passive stellar populations.
However, $\sim$30\% of disks appear to have the colours of passive
stellar populations. These are probably genuinely passive disks or at
least have substantial evolved stellar populations.

6. We also fit the SEDs of our EROs finding slightly better agreement
between the morphological classification and the expected
star-formation activity of such morphological types, than the
\citet{2000MNRAS.317L..17P} classification, i.e.\ most disk-like EROs
exhibit photometric signs of recent star-formation.

7. For a subsample of our EROs we derive SED-fitted photometric
redshift distributions with a median of $z = 1.20\pm0.22$, in
reasonable agreement with spectroscopic redshifts of small samples of
somewhat brighter EROs.

We have highlighted the diversity of Extremely Red Objects, comparing
morphological and photometric classification schemes. We suggest that
the apparent break in ERO counts at K$\sim$19 is due to a rapid rise in
bluer disk EROs at bright magnitudes.  The dominance of this
population declines as it is joined around $K \sim 18.5$ by a more
modestly increasing population of galaxies comprising bulges,
amorphous, merging and increasingly unclassifiable systems.  We have
shown that these blue disk EROs have properties not inconsistent with
them being at the same redshifts as the rest of the other ERO
sub-populations.

The next steps in understanding the nature of the ERO population are:
spectroscopic follow up of a large sample of faint EROs (building on
the initial work of \citealt{2002A&A...381L..68C} at brighter
magnitudes) in order to obtain star formation rates, extinction values
and redshift distributions (particularly to understand the bluer
disk-like EROs); and a detailed comparison with semi-analytic and other
galaxy formation models.

\section*{Acknowledgments}

We thank the referee for useful suggestions which improved the
structure and content of this manuscript.  We thank Carlton Baugh,
Richard Bower, Malcolm Currie, John Lucey, Nigel Metcalfe and Mark
Sullivan for helpful discussions and comments, and also the ST-ECF
staff for providing an excellent service in the WFPC2 associations
archive. We are particularly grateful to Peter Draper for technical
assistance with the {\sc ccdpack} software, and for his expertise with
world coordinate systems.  DGG acknowledges support from the Leverhulme
Trust.  IRS acknowledges support from the Royal Society and Leverhulme
Trust.

\begin{table*}
\caption{Example of our ERO catalogue. Columns show: ERO ID; Field name
(derived from MDS naming); RA; Dec; Total $K$-band magnitude; I-K
colour, MDS Maximum Likelihood classification; concentration index;
visual morphological class; and comments. The full table is available
in the electronic version of the journal.}
\label{table:eros}
\centering
\begin{tabular}{lccccccccp{5cm}}
\hline
ID\# & Field & $\alpha$ (J2000) & $\delta$ (J2000) & $K_{\rm Tot}$ &
$(I - K)$ & MLE & C & Vis & Comments \\
\\
\hline
\hline
001 & U2AY2 & 15:58:48.7 & +42:05:14 &  19.69$\pm$0.14   & 4.60$\pm$ 0.13 & --- & 0.30 & 2 &     \\ 
002 & U2AY2 & 15:58:50.5 & +42:05:03 &  19.63$\pm$0.17   & 4.33$\pm$ 0.13 & bulge & 0.25 & 2 &  early-type? \\ 
003 & U2AY2 & 15:58:46.5 & +42:04:31 &  18.79$\pm$0.09   & 4.20$\pm$ 0.07 & bulge & 0.28 & 6 &  face-on spiral Sbc \\ 
004 & U2H91 & 22:17:33.4 & +00:15:04 &  19.50$\pm$0.14   & 4.72$\pm$ 0.14 & --- & 0.33 & 3 &     \\ 
005 & U2H92 & 13:12:11.4 & +42:44:33 &  19.69$\pm$0.04   & 4.31$\pm$ 0.05 & disk & 0.36 & 3 &    \\ 
006 & U2H92 & 13:12:14.1 & +42:43:56 &  19.67$\pm$0.04   & 4.42$\pm$ 0.04 & bulge & 0.20 & 2 &  little Sa \\ 
007 & U2H92 & 13:12:14.0 & +42:43:06 &  18.98$\pm$0.03   & 4.00$\pm$ 0.03 & d+bgal & 0.24 & 6 &         face-on disk? \\ 
008 & U2IY1 & 03:02:46.2 & +00:13:45 &  19.53$\pm$0.13   & 4.57$\pm$ 0.14 & disk & 0.14 & 9 &   or merger (near chip edge) \\ 
009 & U2IY1 & 03:02:46.5 & +00:13:31 &  19.54$\pm$0.15   & 4.95$\pm$ 0.15 & bulge & 0.24 & 1 &   \\ 
010 & U2IY1 & 03:02:42.0 & +00:13:17 &  19.30$\pm$0.13   & 4.29$\pm$ 0.10 & disk & 0.19 & 6 &   edge-on late-type disk \\ 
... & .... & .. .. ..    &  .. .. .. & ..... & .... & ... & .... & . &
...  \\
... & .... & .. .. ..    &  .. .. .. & ..... & .... & ... & .... & . &
...  \\
\hline
\end{tabular}
\end{table*}

\label{lastpage}
\end{document}